# Fluorescence branching ratios and magnetic tuning of the visible spectrum of SrOH

Duc-Trung Nguyen and Timothy C. Steimle
*School of Molecular Sciences*
*Arizona State University*
*Tempe, AZ 85287-1604*

Ivan Kozyryev
*Department of Physics, Harvard University and Harvard-MIT Center for Ultracold Atoms,*
*Cambridge, MA 02138, USA*

Meng Huang
*Department of Chemistry and Biochemistry*
*The Ohio State University*
*Columbus, OH 43210*

Anne B. McCoy
*Department of Chemistry*
*University of Washington*
*Seattle, WA 98195*

Running Header: "Optical Zeeman effect and Dispersed fluorescence of SrOH"

**Abstract**

The magnetic tuning of the low rotational levels in the $\tilde{X}\,^2\Sigma^+$(0,0,0), $\tilde{A}\,^2\Pi_r$ (0,0,0), and $\tilde{B}\,^2\Sigma^+$(0,0,0) electronic states of strontium hydroxide, SrOH, have been experimentally investigated using high resolution optical Zeeman spectroscopy of a cold molecular beam sample. The observed Zeeman shifts and splittings are successfully modeled using a traditional effective Hamiltonian approach to account for the interaction between the $\tilde{A}\,^2\Pi_r$ and $\tilde{B}\,^2\Sigma^+$ states. The determined magnetic $g$-factors for the $\tilde{X}\,^2\Sigma^+$, $\tilde{A}\,^2\Pi_r$, and $\tilde{B}\,^2\Sigma^+$ states are compared to those predicted by perturbation theory. The dispersed fluorescence resulting from laser excitation of rotationally resolved branch features of the $0^0_0\ \tilde{B}\,^2\Sigma^+ \leftarrow \tilde{X}\,^2\Sigma^+$, $0^0_0\ \tilde{A}\,^2\Pi_{3/2} \leftarrow \tilde{X}\,^2\Sigma^+$ and $0^0_0\ \tilde{A}\,^2\Pi_{1/2} \leftarrow \tilde{X}\,^2\Sigma^+$ transitions have been recorded and analyzed. The measured fluorescence branching ratios are compared with Franck-Condon calculations. The required bending motion wave functions are derived using a discrete variable representation (DVR) method. Implications for laser slowing and magneto-optical trapping experiments for SrOH are described.

Keywords: Strontium hydroxide, Branching ratios, Franck-Condon factors

# I. Introduction

Production of trapped ultracold (<1mK) samples of polyatomic molecules will significantly impact many diverse areas of physics and chemistry. The additional rotational and vibrational degrees of freedom offered by molecules with three and more atoms will open new frontiers in quantum simulation[1], quantum computation[2], tests of fundamental symmetries of nature[3], and studies of inelastic collisions and chemical reactions[4]. Laser cooling and magneto-optical trapping (MOT) allows for production of ultracold and dense atomic samples[5]. However, the absence of closed transitions in molecules significantly hinders optical cycling and laser cooling in even the simplest diatomic species[6]. Recent successful MOTs of SrF[7, 8] and CaF[9-12] has invoked interest in similar isoelectronic triatomic molecules SrOH[13] and CaOH[14]. In the case of SrF the (0,0) $A\,^2\Pi_{1/2}$-$X\,^2\Sigma^+$ transition was employed as the laser cooling transition in order to prevent electronic branching[15]. Additionally, the $B^2\Sigma^+ - X^2\Sigma^+$ transition in CaF was used for laser slowing and longitudinal cooling because of its highly diagonal Franck-Condon factors (FCFs) and small $B\,^2\Sigma^+$ to $A\,^2\Pi$ branching ratio[11].

It has been shown[16] that for these molecules the correct choice of laser polarization critically depends upon the sign and magnitude of the magnetic g-factors and the magnetic hyperfine splitting of the levels associated with the optical transition. In the Hund's case (*a*) limit a $^2\Pi_{1/2}$ state has a negligibly small g-factor and leads to significantly reduced MOT confining forces in the traditional d.c. or type I trapping configuration[16]. The success of the original SrF d.c. MOT is partly due to non-vanishing magnetic tuning in the $A\,^2\Pi_{1/2}$ state caused by mixing with the nearby $B\,^2\Sigma^+$ state. The success of SrF MOT can also be attributed to the fact that the employed $A^2\Pi_{1/2} - X\,^2\Sigma^+$ transition is essentially a Sr$^+$-centered atomic transition and the bonding remains nearly unchanged upon excitation. As a consequence the branching ratios, $b_{v',v''}$, are very diagonal (i.e. $b_{v',v''} \approx 1$ for $v' = v''$ and $b_{v',v''} \approx 0$ for $v' \neq v''$) permitting a multitude of optical excitations before the vibrational state changes. The $0_0^0$ $\tilde{A}^2\Pi_{1/2} - \tilde{X}^2\Sigma^+$ transition of SrOH should have similar characteristics: significant $\tilde{A}^2\Pi_{1/2}$ excited state magnetic tuning due to mixing with $\tilde{B}^2\Sigma^+$ state and nearly diagonal fluorescence. In addition, SrOH has an advantage over SrF because the proton magnetic hyperfine splitting in the $\tilde{A}^2\Pi_{1/2}$ and $\tilde{X}^2\Sigma^+$ states of SrOH is significantly smaller than the $^{19}$F hyperfine splitting in $X^2\Sigma^+$ and $A^2\Pi_{1/2}$ states of SrF. In SrF four magnetic hyperfine substates have to be addressed with four distinct laser frequencies, whereas SrOH should only require two frequencies because the hyperfine splitting is less than the natural linewidth, reducing the experimental complexity. The success of the SrF MOT was also due to use of a helium buffer gas source[17] to generate a slow, cold, molecular beam sample. This source can also generate a cold, slow, intense beam of SrOH[18], which was then used to demonstrate optical cycling and radiative force deflection[13], and most recently Sisyphus laser cooling reducing the transverse translational temperature from 50 mK to approximately 750 μK[19] in one dimension. However, based upon the SrF studies[20], it is expected that scattering in excess of $10^4$ photons per molecule is required to slow the cryogenic buffer-gas beam of SrOH to below the MOT capture velocity, necessitating multiple re-pumping lasers for de-populating excited vibrational levels of the $\tilde{X}^2\Sigma^+$ state.

Here we report on the near natural linewidth limit recording and analysis of the $0_0^0$ $\tilde{B}^2\Sigma^+ - \tilde{X}^2\Sigma^+$, $0_0^0$ $\tilde{A}^2\Pi_{3/2} - \tilde{X}^2\Sigma^+$ and $0_0^0$ $\tilde{A}^2\Pi_{1/2} - \tilde{X}^2\Sigma^+$ transitions of SrOH recorded field-free and in the presence of a static magnetic field. These spectra are analyzed to produce an improved determination of the energy levels, magnetic tuning, and associated spectroscopic parameters in support of proposed MOT measurements. Dispersed fluorescence (DF) spectra arising from the excitation of these bands are also recorded and analyzed to determine the branching ratios, $b_{v',v''}$. A multi-dimensional Franck-Condon factor (FCF) prediction, accounting for the Duschinsky effect and anharmonicity, is performed to predict $b_{v',v''}$. In what follows, mode $v_2$ is the degenerate bending vibration whereas $v_1$ and $v_3$ are approximately the O-H and Sr-O stretching vibrations, respectively.

Spectroscopic studies of SrOH are relatively limited compared to those for SrF. The field-free spectroscopic parameters for the $\tilde{X}\,^2\Sigma^+(0,0,0)$ state are well determined from the analysis of the pure rotational spectrum[21, 22]. A summary of previous studies of excited electronic states can be found in the manuscripts by the Bernath group[23-25]. Particularly relevant to the present investigation is the work by Presunka and Coxon[26-28], involving the Doppler limited excitations of numerous bands in the $\tilde{B}\,^2\Sigma^+ - \tilde{X}\,^2\Sigma^+$ [26] and $\tilde{A}\,^2\Pi_r - \tilde{X}\,^2\Sigma^+$ [27, 28] electronic systems. As part of those studies, DF from various vibrational levels of the $\tilde{A}\,^2\Pi$ and $\tilde{B}\,^2\Sigma^+$ states were recorded and analyzed to produce vibrational spacing for the $\tilde{X}\,^2\Sigma^+$ state. The intensities of the DF spectra were not analyzed to determine branching ratios, which are crucial for determining the feasibility of magneto-optical trapping. Those studies employed a flowing reactor cell to produce a hot sample, which precluded the measurements of numerous low-$J$ branch features, which are relevant to future SrOH MOT experiments.

The most accurate field-free spectroscopic parameters for the $\tilde{X}\,^2\Sigma^+(0,0,0)$ and $\tilde{A}\,^2\Pi(0,0,0)$ states come from the work by Wang et. al[25]. In that study the Doppler limited transition wavenumbers for the $0_0^0\,\tilde{C}\,^2\Pi_r - \tilde{A}\,^2\Pi_r$ band were combined with Doppler limited transition wavenumbers for the $0_0^0\,\tilde{A}\,^2\Pi_r - \tilde{X}\,^2\Sigma^+$ band[29] and the pure rotational transitions[21, 22] in a least squares fit to an effective Hamiltonian model. The spectroscopic parameters for the $B\,^2\Sigma^+(0,0,0)$ state are far less accurately determined and result from the analysis of the Doppler limited spectrum of a high temperature sample, which was performed some time ago[30]. The only previously reported high-resolution optical spectroscopic study of a cold molecular beam sample was that performed as part of an optical-Stark measurement[31] of the $0_0^0\ \tilde{B}\,^2\Sigma^+ - \tilde{X}\,^2\Sigma^+$, $0_0^0\ \tilde{A}\,^2\Pi_{3/2} - \tilde{X}\,^2\Sigma^+$ and $0_0^0\ \tilde{A}\,^2\Pi_{1/2} - \tilde{X}\,^2\Sigma^+$ transitions and the pump/probe microwave double resonance study[32]. Extensive analysis of the field-free electronic spectrum was not performed as part of that study.

As part of a reaction dynamics study, Oberlander and Parson[33] simulated $\tilde{X}\,^2\Sigma^+(v_1,v_2,v_3) \to \tilde{B}\,^2\Sigma^+(v_1,v_2,v_3)$ excitation spectrum using a multi-dimensional FCF prediction following a methodology first presented by Sharp and Rosenstock[34]. In addition to assuming harmonic potentials, that prediction treated the Duschinsky effect (i.e. the transformation for relating the normal coordinates of the initial and final electronic states) by assuming that the same internal coordinates can be adopted for the molecule in the two electronic states involved. Similar assumptions (i.e. harmonic potentials and an approximate Duschinsky effect treatment) were employed to estimate the FCFs for the $\tilde{A}\,^2\Pi_{1/2}(0,0,0) \to \tilde{X}\,^2\Sigma^+(v_1,v_2,v_3)$ emission as part of the optical cycling and radiative force deflection study[13, 35]. In the present study we simulate the $\tilde{B}\,^2\Sigma^+(0,0,0) \to \tilde{X}\,^2\Sigma^+(v_1,v_2,v_3)$ and $\tilde{A}\,^2\Pi_r(0,0,0) \to \tilde{X}\,^2\Sigma^+(v_1,v_2,v_3)$ DF spectra using a more accurate way of handling the Duchinsky effect[36-39] as well as use a numerical integration approach to evaluate the vibrational overlap integrals associated with the bending potentials, which are expected to be anharmonic. Our calculation of the bending mode branching ratios, $b_{v',v''}$, agree with experimental measurements, suggesting that the methodology should be applicable to other linear triatomic alkaline earth molecules.

## II. Experiment

The generation of the pulsed, internally cold ($T^{rot} < 20K$), molecular beam sample via laser ablation/supersonic expansion was identical to that used in the optical Stark measurement.[31] The experimental arrangement for the field-free and optical Zeeman measurements was nearly identical to that used in the recent study of CaF[40]. The molecular beam passes through the poles of a magnetic assemble that consisted of an iron core to which rare-earth magnets are attached. The magnetic field was determined using a commercial Hall-type probe. The excitation source was a single frequency tunable cw-dye laser, whose absolute wavelength is determined to a precision of typically ± 10 MHz by comparison to the sub-Doppler absorption spectrum of $I_2$.[41] The relative position of the cw-dye laser could be determined to an accuracy of approximately ± 3 MHz by monitoring the transmission of an actively stabilized, calibrated, etalon.[42]

The DF spectra were recorded by tuning the cw-dye laser wavelength to be on resonance with an intense branch feature of either the $0_0^0$ $\tilde{B}\,^2\Sigma^+ - \tilde{X}\,^2\Sigma^+$, $0_0^0$ $\tilde{A}\,^2\Pi_{3/2} - \tilde{X}\,^2\Sigma^+$ or $0_0^0$ $\tilde{A}\,^2\Pi_{1/2} - \tilde{X}\,^2\Sigma^+$ bands and accumulating fluorescence signals of a large number (approximately 1000) of pulsed molecular beam samples. As previously described[43], the spectrometer consisted of a 2/3 meter, high efficiency ($f \cong 6.2$), monochromator equipped with a low-dispersion grating (300 lines/mm) and a cooled, gated, ICCD. The camera software bins the vertical column of the two dimensional ICCD arrays to produce fluorescence signal at a given wavelength. At a particular grating angle the ICCD captures a 75 nm spectral window. These segments are subsequently pieced together to produce a continuous DF spectrum across the 580-780 nm range. In the case of the $\tilde{A}\,^2\Pi_r(0,0,0) \to \tilde{X}\,^2\Sigma^+(v_1,v_2,v_3)$ DF measurements the entrance slit of the monochromator was narrowed to 1.0 mm resulting in a spectral resolution of ± 1.2 nm, whereas for the $\tilde{B}\,^2\Sigma^+(0,0,0) \to \tilde{X}\,^2\Sigma^+(v_1,v_2,v_3)$ DF measurements the entrance slit was 2.5 mm resulting in a spectral resolution of ± 3 nm. Wavelength calibration of the resulting DF spectrum was achieved by measuring the emission from an argon pen lamp. The conjoined spectra were flux calibrated for the detection system and background spectra subtracted.

### III. Observation

Excitation spectra in the 14802 cm$^{-1}$ to 14808 cm$^{-1}$ spectral range of the $0_0^0$ $\tilde{A}\,^2\Pi_{3/2} - \tilde{X}\,^2\Sigma^+$ sub-band, the 14540 cm$^{-1}$ to 14545 cm$^{-1}$ spectral range of the $0_0^0$ $\tilde{A}\,^2\Pi_{1/2} - \tilde{X}\,^2\Sigma^+$ sub-band, and the 16373 cm$^{-1}$ to 16381 cm$^{-1}$ spectral range of the $0_0^0$ $\tilde{B}\,^2\Sigma^+ - \tilde{X}\,^2\Sigma^+$ were recorded and assigned. The observed transition wavenumbers, quantum number assignments, and the residuals from the subsequent analysis (see below) for the field-free measurements of the $0_0^0$ $\tilde{A}\,^2\Pi - \tilde{X}\,^2\Sigma^+$ and $0_0^0$ $\tilde{B}\,^2\Sigma^+ - \tilde{X}\,^2\Sigma^+$ bands are presented in Tables 1 and 2, respectively. The spectral features were easily power broadened and exhibited a full width at half maximum width (FWHM) of approximately 50 MHz using an unfocussed laser of approximately 10 mW. Under these low power conditions a slight asymmetry of the lowest-$J$ branch features, presumably due to $\tilde{B}\,^2\Sigma^+(0,0,0)$ and $\tilde{A}\,^2\Pi(0,0,0)$ state proton hyperfine interaction, could be detected.

The DF spectrum of the $\tilde{A}\,^2\Pi_{1/2}(0,0,0)$ state was obtained by cw-laser excitation of the $Q_1(9/2)$ line of the $0_0^0$ $\tilde{A}\,^2\Pi_{1/2} - \tilde{X}\,^2\Sigma^+$ band near 14453.497 cm$^{-1}$ and a typical spectrum is presented in Figure 1. The features near 691.8 nm and 718.0 nm are the $0_0^0$ $\tilde{A}\,^2\Pi_{1/2} \to \tilde{X}\,^2\Sigma^+$ and $3_1^0$ $\tilde{A}\,^2\Pi_{1/2} \to \tilde{X}\,^2\Sigma^+$ transitions, respectively. To account for any contribution from Sr $^3P_1 \to\ ^1S$ emission ($\lambda_{air}$ = 689.26 nm) in the observed 691.8 nm feature in Figure 1, a background spectrum with the dye laser blocked was recorded and subtracted. To account for any laser scatter, a background spectrum of only the dye laser signal in the absence of a molecular beam was also subtracted. The DF spectrum of the $\tilde{A}^2\Pi_{3/2}(0,0,0)$ states was obtained by cw-laser excitation of the $R_2(7/2)$ line of the $0_0^0$ $\tilde{A}^2\Pi_{3/2} - \tilde{X}^2\Sigma^+$ band near 14806.952 cm$^{-1}$ and a typical spectrum is presented in Figure 2. The features near 675.4 nm and 700.2 nm are the $0_0^0$ $\tilde{A}^2\Pi_{3/2} \to \tilde{X}^2\Sigma^+$ and $3_1^0$ $\tilde{A}^2\Pi_{3/2} \to \tilde{X}^2\Sigma^+$ emissions, respectively. As in the case of the $\tilde{A}^2\Pi_{1/2}$ experiments, to account for any contribution from Sr $^3P_1 \to {}^1S$ emission ($\lambda_{air}$ = 689.26 nm) in the observed 691.8 nm feature, a background spectrum with the dye laser blocked was recorded and subtracted. Excitation of the $R_1(13/2)$ line of the $0_0^0$ $\tilde{B}^2\Sigma^+ - \tilde{X}^2\Sigma^+$ band at 16380.648 cm$^{-1}$ was used for the DF measurement of the $\tilde{B}^2\Sigma^+(0,0,0)$ state, with a typical spectrum presented in Figure 3. In this case the slits of the monochromator were opened slightly to enhance the detection of the non-diagonal peak. The features near 610.5 nm and 630.7 nm are the $0_0^0$ $\tilde{B}^2\Sigma^+ \to \tilde{X}^2\Sigma^+$ and $3_1^0$ $\tilde{B}^2\Sigma^+ \to \tilde{X}^2\Sigma^+$ emissions, respectively. The spectral feature near 690 nm is the Sr $^3P_1 \to {}^1S$ emission from metastable atoms generated in the laser ablation source. Unlike the DF spectra associated with emission from the $\tilde{A}^2\Pi_{1/2}$ and $\tilde{A}^2\Pi_{3/2}$ levels, subtracting a background spectrum with the dye laser blocked was not performed because the Sr $^3P_1 \to {}^1S$ emission ($\lambda_{air}$ = 689.26 nm) does not overlap with SrOH emission features. Also indicated by the arrows in Figures 1-3 are the expected locations of emission to the $\tilde{X}^2\Sigma^+(0,1^1,0)$ and $\tilde{X}^2\Sigma^+(0,2^0,0)$ states. Emission to the $\tilde{X}^2\Sigma^+(1,0,0)$ level, which is expected to be at approximately[44] 3778 cm$^{-1}$ to the red of the excitation, is outside the operating range of the ICCD. The $\tilde{B}^2\Sigma^+(0,0,0) \to \tilde{X}^2\Sigma^+(1,0,0)$ and $\tilde{A}^2\Pi(0,0,0) \to \tilde{X}^2\Sigma^+(1,0,0)$ emissions were not detected in the previous flowing reactor DF spectra, indicating a small branching ratio for decays to the excited O-H stretching vibrations, consistent with our calculations (see below).

The observed field-free spectrum for the low-$J$ $R_{12}$ and $Q_1$ branch features of the $0_0^0$ $\tilde{A}\,^2\Pi_{1/2} - \tilde{X}\,^2\Sigma^+$ band in the 14452.6 cm$^{-1}$ to 14453.1 cm$^{-1}$ spectral range, and the low-$J$ $R_2$ and $^RQ_{21}$ branch features of the $0_0^0$ $\tilde{A}\,^2\Pi_{3/2} - \tilde{X}\,^2\Sigma^+$ band in the 14805.9 cm$^{-1}$ to 14806.7 cm$^{-1}$ spectral range, are presented in Figures 4 and 5, respectively. Also presented are the observed and predicted 477 G Zeeman spectra with parallel ($\Delta M_J = 0$) and perpendicular ($\Delta M_J = \pm 1$) orientations relative to linear polarized laser. The $R_{12}(J'')$ and $Q_1(J''+1/2)$ branch features and the $R_2$ and $^RQ_{21}$ branch features are separated by the electron spin-rotation splitting ("ρ-doubling") of the $\tilde{X}\,^2\Sigma^+(0,0,0)$ state. The observed field-free spectrum for the low-$J$ $R_1$ and $^RQ_{12}$ branch features of the $0_0^0$ $\tilde{B}\,^2\Sigma^+ \rightarrow \tilde{X}\,^2\Sigma^+$ band in the 16377.9 cm$^{-1}$ to 16378.4 cm$^{-1}$ spectral range along with the observed and predicted 913 G Zeeman spectra are presented in Figure 6. The observed field-free transition wavenumbers, assignments, and observed-calculated residuals are presented in the Tables 1 and 2, respectively. The Zeeman induced shifts, spectral assignments and the observed-calculated residuals for the $0_0^0$ $\tilde{A}\,^2\Pi_{1/2} - \tilde{X}\,^2\Sigma^+$ $0_0^0$ $\tilde{A}\,^2\Pi_{3/2} - \tilde{X}\,^2\Sigma^+$, and $0_0^0$ $\tilde{B}\,^2\Sigma^+ - \tilde{X}\,^2\Sigma^+$ bands, respectively, are provided in Tables S1, S2 and S3 of the Supplemental Material.

### IV. Analysis

#### a. Field-free Spectra

A direct fit to the measured field-free transition frequencies was performed. The effective Hamiltonian operator for the $\tilde{X}\,^2\Sigma^+(0,0,0)$ and $\tilde{B}\,^2\Sigma^+(0,0,0)$ states included the origin, rotation, its centrifugal distortion, and spin-rotation terms[45]:

$$\mathbf{H}^{eff}(^2\Sigma) = T_v + B\mathbf{N}^2 - D\mathbf{N}^4 + \gamma \mathbf{N}\cdot\mathbf{S} \ . \qquad 1)$$

The proton hyperfine contributions were excluded because only a very slight asymmetry (< 10 MHz) of the spectral lineshape was observed. The effective Hamiltonian operator for the $\tilde{A}^2\Pi_r(0,0,0)$ state including the origin, spin-orbit interaction and associated centrifugal distortion correction, rotation and associated centrifugal distortion correction, and the Λ-doubling interaction terms[45]:

$$\mathbf{H}^{\text{eff}}(\tilde{A}^2\Pi) = T_v + A L_z S_z + A_D \frac{1}{2}\left[\mathbf{N}^2, L_z S_z\right]^+ + B\mathbf{N}^2 - D\mathbf{N}^4 +$$
$$\frac{1}{2}(p+2q)(e^{2i\varphi}S_-J_- + e^{-2i\varphi}S_+J_+) - \frac{1}{2}q(e^{2i\varphi}J_-^2 + e^{-2i\varphi}J_+^2)$$  (2)

In Eq. 2 $J_\pm$ are the shift operators of the total angular momentum in the absence of the nuclear spin, [ ]$^+$ is the anti-commutator, and φ is the azimuthal angle of the electron. The field-free eigenvalues and eigenvectors for the $\tilde{X}^2\Sigma^+(0,0,0)$ and $\tilde{B}^2\Sigma^+(0,0,0)$ states were obtained by diagonalization of a 2×2 matrix representation constructed using a Hund's case (*a*), non-parity, basis set. For the $\tilde{A}^2\Pi_r$ state a 4 × 4 matrix representation was constructed and diagonalized to produce the eigenvalues and eigenvectors. A nonlinear least-squares fitting procedure program was written that used as input the observed transition wavenumbers of Tables 1 and 2 and optimize the parameters for the $\tilde{A}^2\Pi_r(0,0,0)$ and $\tilde{B}^2\Sigma^+(0,0,0)$ states separately. The rotational, *B* (0.249199914 cm$^{-1}$), centrifugal distortion correction to rotation, *D* (2.17437×10$^{-7}$ cm$^{-1}$), spin-rotational, γ (2.4274803×10$^{-3}$ cm$^{-1}$), parameters for the $\tilde{X}^2\Sigma^+(0,0,0)$ state and the $A_D$ (1.331×10$^{-3}$ cm$^{-1}$), *D* (2.1685×10$^{-7}$ cm$^{-1}$), and *q* (-1.528×10$^{-4}$ cm$^{-1}$) parameters for the $\tilde{A}^2\Pi_r(0,0,0)$ state were held fixed to those determined from the combined analysis of the millimeter wave and optical spectra[25]. For the $\tilde{B}^2\Sigma^+(0,0,0)$ state, *D* (2.15×10$^{-7}$ cm$^{-1}$) was held fixed to the value that determined from the analysis of the high-temperature excitation spectrum[30]. The determined field-free spectroscopic parameters for the $\tilde{A}^2\Pi_r(0,0,0)$ and $\tilde{B}^2\Sigma^+(0,0,0)$ states are presented in Table 3 as are the previously determined values. The standard deviation of the fits for the 82 measured transition wavenumbers of the $0_0^0$ $\tilde{A}\,^2\Pi_r - \tilde{X}\,^2\Sigma^+$ band (Table 1), and the 40 measured transition wavenumbers of the $0_0^0$ $\tilde{B}\,^2\Sigma^+ - \tilde{X}\,^2\Sigma^+$ band (Table 2), are $4.4 \times 10^{-4}$ cm$^{-1}$ and $8.2 \times 10^{-4}$ cm$^{-1}$, respectively, which are commensurate with the estimated measurement error.

### b. Zeeman Spectra

The effective Zeeman Hamiltonian operator was taken as[45]:

$$H_{eff}^{Zee} = g_S \mu_B \hat{\mathbf{S}} \cdot \hat{\mathbf{B}} + g'_L \mu_B \hat{\mathbf{S}} \cdot \hat{\mathbf{B}} + g_l \mu_B (\hat{S}_x \hat{B}_x + \hat{S}_y \hat{B}_y) + g'_l \mu_B (e^{-2i\phi} \hat{S}_+ \hat{B}_+ + e^{+2i\phi} \hat{S}_- \hat{B}_-) \qquad 3)$$

where $\hat{S}_x$ and $\hat{B}_x$ refer to the *x*-axis molecule-fixed components of the electronic spin angular momentum and magnetic field, respectively, and $\phi$ is the azimuthal angle of the electronic coordinates. Modeling the Zeeman effect in a $^2\Sigma^+$ state requires varying $g_l$, and to a lesser extent $g_S$, while modeling the Zeeman effect in a state with non-zero orbital angular momentum requires varying $g'_L$, $g_l$, $g'_l$, and possibly $g_S$. The eigenvalues and eigenvectors of the $\tilde{X}\,^2\Sigma^+(0,0,0)$ and $\tilde{B}\,^2\Sigma^+(0,0,0)$ states were modeled by diagonalization of the 12×12 matrix constructed using the Hund's case (*a*) basis set functions for the $J \leq 11/2$ levels. Similarly, the eigenvalues and eigenvectors of the $\tilde{A}\,^2\Pi_r(0,0,0)$ state were modeled by diagonalization of a 24×24 Hund's case (*a*) representation for the $J \leq 11/2$ levels. This truncation was sufficient to reproduce the experimental accuracy (± 30 MHz) of the $0_0^0$ $\tilde{B}\,^2\Sigma^+ - \tilde{X}\,^2\Sigma^+$, $0_0^0$ $\tilde{A}\,^2\Pi_{3/2} - \tilde{X}\,^2\Sigma^+$ and $0_0^0$ $\tilde{A}\,^2\Pi_{1/2} \leftarrow \tilde{X}\,^2\Sigma^+$ Zeeman data set, which included maximum *J* values of 9/2.

Least squares fits of the magnetic field induced shifts using various combinations of the Zeeman parameters were attempted. In the end satisfactory fits were obtained by constraining $g_S$, and $g'_L$ to their nominal values of 2.0023 and unity, and constraining $g_l$ for the $\tilde{X}\,^2\Sigma^+(0,0,0)$ state to the value of -4.87x10$^{-3}$ predicted by the Curl relationship[45] ($g_l \approx -\gamma/2B$). The data was insensitive to the parity dependent term, $g'$, and it was constrained to zero. The optimized $g_l$ parameter for the $\tilde{A}^2\Pi_r(0,0,0)$ and $\tilde{B}^2\Sigma^+(0,0,0)$ states and associated errors are presented in Table 3. The standard deviation of the fits for $0_0^0$ $\tilde{A}^2\Pi_r - \tilde{X}^2\Sigma^+$ band and the $0_0^0$ $\tilde{B}^2\Sigma^+ - \tilde{X}^2\Sigma^+$ band are 28 MHz and 31 MHz, respectively, which are commensurate with the estimated measurement error of the Zeeman shifts.

### c. Dispersed Fluorescence Spectra

The DF spectra were corrected for the wavelength response of the spectrometer. The ratios of the integrated areas of the corrected DF features were used to determine the branching ratios given in Table 4. The branching ratios were taken as the ratios of the observed integrated intensities, which are assumed be proportional to the ratio of FCFs, and transition frequency cubed[46]:

$$b_{v',v''} \equiv \frac{I_{iv',fv''}}{\sum_{fv''} I_{iv',fv''}} \approx \frac{\mathrm{FCF}_{iv',fv''} \times v^3_{iv',fv''}}{\sum_{fv''} \mathrm{FCF}_{iv',fv''} \times v^3_{iv',fv''}} \qquad . \qquad 4)$$

Also presented in Table 4 are the branching ratios from the predicted Franck-Condon factors (see below).

### d. Modeling the spectra

The quantum number assignment of the Zeeman spectra was greatly assisted by spectral intensity simulations. This was achieved by constructing the Hund's case (*a*) transition moment matrices for the $0_0^0$ $\tilde{A}^2\Pi_r - \tilde{X}^2\Sigma^+$ and $0_0^0$ $\tilde{B}^2\Sigma^+ - \tilde{X}^2\Sigma^+$ bands. The transition moment was obtained by cross multiplication of the transition moment matrix by the Hund's case (*a*) eigenvectors. The transition moment was squared, multiplied by a Boltzmann factor commensurate with a rotational temperature of 10 K and used in conjunction with a Lorentzian linewidth of 30 MHz full width at half maximum to predict each spectral feature. The predicted spectra, such as those given in Figs. 4-6, were obtained by co-adding the individual spectral feature components.

## V. Discussion

The determined field-free spectroscopic parameters for the $\tilde{B}^2\Sigma^+(0,0,0)$ state (see Table 3) are significantly improved from the previous values. The determined field-free spectroscopic parameters for the $\tilde{A}^2\Pi_r(0,0,0)$ state are comparable to the previous values because although the current data set is much smaller the measured transition wavenumbers are substantially more precise (0.0001 cm$^{-1}$ vs 0.003 cm$^{-1}$). The improved fine structure constants for the $\tilde{A}^2\Pi_r(0,0,0)$ and $\tilde{B}^2\Sigma^+(0,0,0)$ states can be used to predict the transition wavenumbers associated with planned laser cooling and magneto-optical trapping experiments. The very small standard deviations from the field-free fits of 0.00091 cm$^{-1}$ and 0.00049 cm$^{-1}$ for the $0_0^0$ $\tilde{A}^2\Pi_r - \tilde{X}^2\Sigma^+$ and $0_0^0$ $\tilde{B}^2\Sigma^+ - \tilde{X}^2\Sigma^+$ bands, respectively, demonstrates that the effective Hamiltonian approach can accurately model the low rotational lines of these bands, and more generally for other linear alkaline earth monovalent triatomic molecules.

The magnetic tuning of the rotational lines associated with magneto-optical trapping of SrOH have been measured and successfully modelled with an effective Hamiltonian approach. The determined g'$_l$ and g$_l$ of -0.267(6) and 0.283(16) for the magnetic g-factors given in Table 3 for the $\tilde{A}^2\Pi_r(0,0,0)$ and $\tilde{B}^2\Sigma^+(0,0,0)$ states, respectively, having nearly the same magnitude (within the uncertainty) but opposite sign indicating a high degree of mixing between these two states. The Curl-type relationships[45] $g'_l(\tilde{A}^2\Pi) \approx p/2B$ and $g_l(\tilde{B}^2\Sigma^+) \approx -\gamma/2B$ predict values of

-0.2824 and 0.2827, respectively, which is good agreement with the determined values of -0.267(6) and 0.283(16).

The DF spectra resulting from excitation to the $\tilde{A}\,^2\Pi_r(0,0,0)$ and $\tilde{B}\,^2\Sigma^+(0,0,0)$ states were modeled by assuming that the relative intensities are proportional to the product of the two-dimensional FCF of the $\sigma$-symmetry stretching modes ($v_1$ and $v_3$) and a FCF for the one-dimensional $\pi$-symmetry bending mode ($v_2$):

$$\text{FCF} = \left|\langle v_1'' v_3'' | 0,0 \rangle\right|^2 \left|\langle v_2'' | 0 \rangle\right|^2 \qquad \qquad 5)$$

This assumes that vibronic coupling due to Renner-Teller and spin-orbit interactions[47, 48] is negligible and the total wave function can be written as the product of an electronic and vibrational wave function, which is a reasonable assumption for the vibronic states considered here. The previous prediction[13] of FCFs followed the procedure of Sharp and Rosenstock[34] to evaluate the two dimensional integrals of Eq. 5. Here an alternative[49] closed-form formula for the two-dimensional FCFs for the stretching modes is employed. Both formulations assume a harmonic stretching motion. The Sharp and Rosenstock[34] approach also assumes that the same internal coordinates of the ground and excited electronic state can be used for treating the Duschinsky effect (i.e. mode mixing). A more accurate method of handling the Duschinsky effect, which is used here, employs Cartesian coordinates common to both electronic states[36, 38, 39]. This approach requires relating the normal coordinates of the $\tilde{X}\,^2\Sigma^+$ state, $\mathbf{Q}(\tilde{X}\,^2\Sigma^+)$, to those of the $\tilde{A}\,^2\Pi$ and $\tilde{B}\,^2\Sigma^+$ states, $\mathbf{Q}(\tilde{A}\,^2\Pi, \tilde{B}\,^2\Sigma^+)$:

$$\mathbf{Q}(\tilde{X}\,^2\Sigma^+) = \mathbf{J}\mathbf{Q}(\tilde{A}\,^2\Pi, \tilde{B}\,^2\Sigma^+) + \mathbf{D} \qquad \qquad 6)$$

In Eq. 6, **D** is the vector of geometry displacements given in terms of the normal coordinates of the ground state and **J** is the Duschinsky rotation matrix which for the $\tilde{A}^2\Pi_r - \tilde{X}^2\Sigma^+$ transition are given by:

$$\mathbf{J} = (\mathbf{L}^{-1}(\tilde{A}^2\Pi_r) \cdot \mathbf{B}(\tilde{A}^2\Pi_r)) \cdot \mathbf{M}^{-1} \cdot ((\mathbf{L}^{-1}(X^2\Sigma^+) \cdot \mathbf{B}(X^2\Sigma^+))^T \qquad 7)$$

and

$$\mathbf{D} = (\mathbf{L}^{-1}(\tilde{A}^2\Pi_r) \cdot \mathbf{B}(\tilde{A}^2\Pi_r)) \cdot (\mathbf{R}_{eq}(\tilde{A}^2\Pi_r) - \mathbf{R}_{eq}(\tilde{X}^2\Sigma^+)) \qquad 8)$$

In Eqs. 7 and 8 the **L** matrices relate the internal displacement coordinated to the normal coordinates and **R$_{eq}$** the vectors of equilibrium Cartesian coordinates in a center of mass where the atoms lie along the z-axis. For the linear-to-linear transition studied here the 2×2 rotation matrix associated with the two σ-type stretching modes is a unit matrix. The derivation of the displacement vectors for the σ-type stretching modes from a normal coordinate analyses of the $\tilde{X}^2\Sigma^+$, $\tilde{A}^2\Pi$, and $\tilde{B}^2\Sigma^+$ states is presented in Appendix A. Also presented in Appendix A is the derivation of the **G** matrix element for the bending mode, which is required for the one dimensional overlap integral. The two dimensional FCFs $|\langle \tilde{A}^2\Pi,0,0|\tilde{X}^2\Sigma^+,0,0\rangle|^2$, $|\langle \tilde{A}^2\Pi,0,0|\tilde{X}^2\Sigma^+,0,1\rangle|^2$ $|\langle \tilde{A}^2\Pi,0,0|\tilde{X}^2\Sigma^+,0,2\rangle|^2$, $|\langle \tilde{B}^2\Sigma^+,0,0|\tilde{X}^2\Sigma^+,0,0\rangle|^2$, $|\langle \tilde{B}^2\Sigma^+,0,0|\tilde{X}^2\Sigma^+,0,1\rangle|^2$ and, $|\langle \tilde{B}^2\Sigma^+,0,0|\tilde{X}^2\Sigma^+,0,2\rangle|^2$, are presented in Table 4. All other two dimensional FCFs for the σ-type stretching modes involving emission from the $\tilde{A}^2\Pi_r(0,0,0)$ and $\tilde{B}^2\Sigma^+(0,0,0)$ states are less than $1\times10^{-4}$.

The bending potentials for the $\tilde{X}^2\Sigma^+$, $\tilde{A}^2\Pi_r$, and $\tilde{B}^2\Sigma^+$ states of SrOH are expected to be anharmonic, similar to what has been observed for CaOH and MgOH [50-53]. To evaluate the one dimensional bending FCF of Eq. 5 a numerical approach using a two-dimensional (2D) discrete variable representation (DVR) technique[54-57] was employed and is described in Appendix B. The DVR approach is based upon a grid-point representation and replaces the problem of integral evaluation associated with the vibronic Schrödinger equation with a numerical summation of the matrix elements of the kinetic and potential energy operators at the grid points. This approach is particularly useful for the bending FCFs because it does not require an analytic expression for the potential. As applied here the representation for the potential energy operator was obtained from an electronic structure calculation, using the ORCA[58] suite of program. The properties of the $\tilde{X}\,^2\Sigma^+$ state was calculated using Density Functional Theory (DFT) implemented the BP86 functional and def2-QZVPP basis set, while those for the $\tilde{A}\,^2\Pi$ and $\tilde{B}\,^2\Sigma^+$ states were calculated using Time Dependent (TD)-DFT with the same functional and basis set. The potential energies for the $\tilde{X}\,^2\Sigma^+$, $\tilde{A}\,^2\Pi_r$, and $\tilde{B}\,^2\Sigma^+$ states as a function of bond angles ranging from 180° to 90° were calculated in 1° steps. The potential is symmetrical about 180° giving a total of 181 energies for each electronic state. At each angle the Sr-O and O-H bond lengths were optimized. Details of the procedure and the determined energies are given in the Supplemental Material. The representation for the kinetic energy operator was obtained from analytical expressions of the G-matrix elements[59] for the bending normal coordinate (see Appendix B). The bending FCFs $\left|\left\langle \tilde{A}\,^2\Pi,0 \middle| \tilde{X}\,^2\Sigma^+,0 \right\rangle\right|^2$, $\left|\left\langle \tilde{A}\,^2\Pi,0 \middle| \tilde{X}\,^2\Sigma^+,2 \right\rangle\right|^2$, $\left|\left\langle \tilde{B}\,^2\Sigma^+,0 \middle| \tilde{X}\,^2\Sigma^+,0 \right\rangle\right|^2$, and, $\left|\left\langle \tilde{B}\,^2\Sigma^+,0 \middle| \tilde{X}\,^2\Sigma^+,2 \right\rangle\right|^2$, obtained from the 2D-DVR calculations are presented in Table 4. All other bending FCFs for emission from the $\tilde{A}\,^2\Pi_r(0,0,0)$ and $\tilde{B}\,^2\Sigma^+(0,0,0)$ states are less than $1\times10^{-4}$.

The predicted branching ratios, $b_{v',v''}$, as calculated using Eq. 4, are also presented in Table 4 and compared with observations. The predictions are in excellent agreement for the six ($0_0^0$, $3_1^0$ $\tilde{A}^2\Pi_{1/2,3/2} - \tilde{X}^2\Sigma^+$ and $0_0^0$, $3_1^0$ $\tilde{B}^2\Sigma^+ - \tilde{X}^2\Sigma^+$ bands) experimentally determined $b_{v',v''}$ values, and reflect the near diagonal nature of the transitions. While the experimental sensitivity of the current measurement allowed us to only observe the dominant off-diagonal $3_1^0$ $\tilde{A}^2\Pi_r - \tilde{X}^2\Sigma^+$ and $3_1^0$ $\tilde{B}^2\Sigma^+ - \tilde{X}^2\Sigma^+$ transitions, we have used the molecular loss data from the SrOH laser cooling experiment[19] (see Appendix C) to test the accuracy of our predictions for smaller vibrational $b_{v',v''}$ values. From the analysis in Appendix C, the measured $3_2^0$ $\tilde{A}^2\Pi_{1/2} - \tilde{X}^2\Sigma^+$ and $2_2^0$ $\tilde{A}^2\Pi_{1/2} - \tilde{X}^2\Sigma^+$ branching ratios are $2.2^{+0.9}_{-0.7} \times 10^{-3}$ and $4^{+2}_{-1} \times 10^{-4}$, respectively, which are in close agreement with the predicted values (see Table 4) of $1.5 \times 10^{-3}$ and $1.0 \times 10^{-4}$, respectively, reaffirming the accuracy of our calculations.

Finally, it is somewhat surprising that the observed (<0.007) and predicted $b_{v',v''}$ (<0.002) values for the symmetry allowed $2_2^0$ $\tilde{A}^2\Pi_r - \tilde{X}^2\Sigma^+$ and $2_2^0$ $\tilde{B}^2\Sigma^+ - \tilde{X}^2\Sigma^+$ transitions are so small given the expected anharmonic nature of the bending potentials (see Supplemental Material). Although not observed here, both the symmetry allowed $\tilde{A}^2\Pi(0,0,0) \to \tilde{X}^2\Sigma^+(0,2^0,0)$ emission and symmetry forbidden $\tilde{A}^2\Pi(0,0,0) \to \tilde{X}^2\Sigma^+(0,1^1,0)$ emission have been seen in the DF spectra [27-29] of a high temperature flowing reactor source. Although no estimates of the relative intensities of the weak $\tilde{A}^2\Pi(0,0,0) \to \tilde{X}^2\Sigma^+(0,2^0,0)$ emission to that of intense $\tilde{A}^2\Pi(0,0,0) \to \tilde{X}^2\Sigma^+(0,0,0)$ emission was given, it was estimated[29] that the nominally symmetry forbidden $\tilde{A}^2\Pi_{1/2}(0,0,0) \to \tilde{X}^2\Sigma^+(0,1^1,0)$ emission was 1/8 that of the symmetry allowed $\tilde{A}^2\Pi_{1/2}(0,0,0) \to \tilde{X}^2\Sigma^+(0,2^0,0)$ emission. Similarly, the $\tilde{B}^2\Sigma^+(0,0,0) \to \tilde{X}^2\Sigma^+(0,1^1,0)$ emission has been seen in the DF spectrum of a rotationally hot sample[30] where it was estimated to be a factor of 1000 less intense than the $\tilde{B}\,^2\Sigma^+(0,0,0) \rightarrow \tilde{X}\,^2\Sigma^+(0,0,0)$ emission. The observations here demonstrate these emissions are much weaker for the low rotational levels relevant to the proposed SrOH MOT experiment[35]. Our results are also consistent with the previous *ab initio* calculation for CaOH[14] and, provide strong experimental evidence for the feasibility of direct laser slowing and magneto-optical trapping of SrOH with four distinct laser wavelengths using either $\tilde{A}\,^2\Pi_r - \tilde{X}\,^2\Sigma^+$ or $\tilde{B}\,^2\Sigma^+ - \tilde{X}\,^2\Sigma^+$ cycling transitions.

## VI. Summary

The field-free frequencies, magnetic tuning of the $0_0^0$ $\tilde{B}\,^2\Sigma^+ - \tilde{X}\,^2\Sigma^+$, $0_0^0$ $\tilde{A}\,^2\Pi_{3/2} - \tilde{X}\,^2\Sigma^+$ and $0_0^0$ $\tilde{A}\,^2\Pi_{1/2} - \tilde{X}\,^2\Sigma^+$ transitions, and the fluorescence branching ratios of the $\tilde{A}\,^2\Pi_r$ (0,0,0), and $\tilde{B}\,^2\Sigma^+$(0,0,0) electronic states determined here should be useful for the continuing effort[13, 18, 19] towards direct laser slowing and magneto-optical trapping of SrOH. It is demonstrated that the proton hyperfine splitting of the optical transitions relevant to generation of MOT sample are less than 30 MHz and that like SrF, the low-rotational features of $\tilde{A}\,^2\Pi_{1/2}$ have a significant magnetic moment due to spin-orbit interaction with the $\tilde{B}\,^2\Sigma^+$ state, which will enable production of a d.c. MOT of SrOH. The predicted FCFs, which excluded the effects of ro-vibronic coupling, are in good agreement with the observed branching ratios. The DVR method for prediction the bending mode eigenvalues and eigenvectors needed for the FCF predictions of linear triatomic molecules has been presented.

Experimental and theoretical results presented here indicate that laser slowing of the cryogenic buffer-gas beam and subsequent loading of a d.c. or r.f. MOT can be achieved with at most four re-pumping lasers, which is comparable to the number of lasers used in the MOTs of SrF[7, 8] and CaF[9-12]. Thus, a technically straightforward path towards direct production of ultracold polyatomic molecules via laser cooling has been confirmed. We anticipate that diagonal FCF arrays should exist for the similar linear triatomic molecules CaOH, BaOH, and YbOH, which we are now investigating. Additionally, other strontium monovalent polyatomic radicals such as strontium monoalkoxides have recently been considered for laser cooling and trapping[60] and future experimental and theoretical analysis similar to the present work on SrOH will be necessary in order to specify the optimal conditions for magneto-optical trapping of complex polyatomic species.

## Acknowledgements

Support by grants from the National Science Foundation (TN and TCS: CSDM-A; CHE-1265885; ABM: CTC; CHE-1619660; IK PHY-1505961) and from the Graduate School at The Ohio State University (MH: Presidential Fellowship) is gratefully acknowledged. The authors thank Prof. Andrew Chizmeshya (School of Molecular Sciences, Arizona State University) for his helpful comments and Prof. John M. Doyle (Department of Physics, Harvard University) for suggesting this project.

**Appendix A:  The two dimensional stretching Franck-Condon factors.**

Normal coordinate analyses using the **GF** matrix approach is well documented by Wilson Decius & Cross[61]. The 4×4 **GF** matrix is block diagonal consisting of a 2×2 GF matrix associated with σ-symmetry stretching modes and two 1×1 matrices for the two orthogonal π-symmetry bending modes. Selecting the z-axis to coincide with the H-O-Sr bond and following the procedure described in Ref. 61 gives for the **B** matrix:

$$\mathbf{B} = \begin{vmatrix} & x_H & y_H & z_H & x_O & y_O & z_O & x_{Sr} & y_{Sr} & z_{Sr} \\ \Delta r_{OH} & 0 & 0 & 1 & 0 & 0 & -1 & 0 & 0 & 0 \\ \Delta r_{SrO} & 0 & 0 & 0 & 0 & 0 & 1 & 0 & 0 & -1 \\ \Delta \theta_{SrOH}^{xz} & a & 0 & 0 & b & 0 & 0 & c & 0 & 0 \\ \Delta \theta_{SrOH}^{yz} & 0 & a & 0 & 0 & b & 0 & 0 & c & 0 \end{vmatrix} , \quad (A1)$$

where $a = -r_{OH}^{-1}$, $b = r_{OH}^{-1} + r_{SrO}^{-1}$, and $c = r_{SrO}^{-1}$. It follows that the **G** matrices are:

$$\mathbf{G} = \begin{vmatrix} & \Delta r_{OH} & \Delta r_{SrO} & \Delta \phi_{SrOH}^{xz} & \Delta \phi_{SrOH}^{yz} \\ \Delta r_{OH} & a & c & 0 & 0 \\ \Delta r_{SrO} & c & b & 0 & 0 \\ \Delta \theta_{SrOH}^{xz} & 0 & 0 & d & 0 \\ \Delta \theta_{SrOH}^{yz} & 0 & 0 & 0 & d \end{vmatrix} \quad (A2)$$

where $a = M_O^{-1} + M_H^{-1}$, $b = M_O^{-1} + M_{Sr}^{-1}$, $c = -M_O^{-1}$, and $d = (M_H r_{OH}^2)^{-1} + (M_{Sr} r_{SrO}^2)^{-1} + (M_O)^{-1}(r_{OH}^{-2} + r_{SrO}^{-2} + 2/(r_{OH} r_{SrO}))$. The required force constants and bond distances for construction of the **GF** matrices are collected in Table A1. The force constants of Table A1 were obtained using primarily experimental data with the exception of the stretch-stretch coupling constant, $f_{13}$, which was set to the value predicted for the $\tilde{X}\,^2\Sigma^+$ state of CaOH[52]. The predictions are least sensitive to this force constant. The O-H and Sr-O stretch force constants, $f_{11}$ and $f_{33}$, were obtained by non-linear least squares fitting the $\omega_1$(SrOH), $\omega_3$ (SrOH), and $\omega_3$ (SrOD) frequencies given in Table A1. For this purpose the 2×2 **GF** sub-matrices associated with σ-symmetry stretching modes were constructed using initial guesses for $f_{11}$ and $f_{33}$ and diagonalized to produce predicted harmonic vibrational frequencies. The difference between the observed and predicted frequencies was minimized in non-linear least squares procedure. The statistical errors associated with fitting the three frequencies $\omega_1$(SrOH), $\omega_3$ (SrOH), and $\omega_3$ (SrOD) to the two parameters $f_{11}$ and $f_{33}$ are also given in Table A1. Similarly, the $f_{22}$ force constants for the bending mode were obtained from the 1×1 sub-matrices.

Table A1. Force constant and bond lengths used in the normal coordinate analysis.

| Property | $\tilde{X}\,^2\Sigma^+$ | $\tilde{A}\,^2\Pi$ | $\tilde{B}\,^2\Sigma^+$ |
|---|---|---|---|
| $f_{11}$[a] | 7.9402±0.0417 | 7.9398±0.0562 | 7.9399±0.0438 |
| $f_{22}$ | 2.2859±0.0651 | 2.3943±0.0878 | 2.3626±0.0684 |
| $f_{33}$ | 0.0590 | 0.0648 | 0.0711 |
| $f_{13}$[b] | 0.0677 | 0.0677 | 0.0677 |
| $r_{Sr-O}$(Å)[c] | 2.111 | 2.091 | 2.098 |
| $r_{O-H}$(Å)[c] | 0.922 | 0.922 | 0.921 |
| $\omega_1$(O-H stretch)[d] | 3778 | 3778 | 3778 |
| $\omega_2$(Sr-O-H bend) | 363.7[e] | 381.4[f] | 399.8[e] |
| $\omega_3$(Sr-O stretch) | 526.9[g] (510.0)[h] | 542.6[e] (516.0)[h] | 536.3[33](516.0)[h] |

a) $f_{11}$= O-H stretch, $f_{33}$ = Sr-O stretch, $f_{22}$ = Sr-O-H bend, $f_{13}$ = stretch-stretch coupling. Units: $f_{11}, f_{33}$, and $f_{13}$ in mdyne/Å; $f_{22}$ mdyne/(Å radian).
b) Constrained to the predicted value for CaOH[52].
c) Ref. 24.
d) Constrained to the experimentally determined value for the $\tilde{X}\,^2\Sigma^+$ state of CaOH[44].
e) Ref. 26
f) Ref. 27
g) Ref. 28
h) Ref. 30
i) Ref. 33

The **L** matrices that appear in Eqs. 7 and 8 are related to the eigenvectors of the **GF** matrices, **V**, by a normalization matrix **N**:

$$\mathbf{L}=\mathbf{V}\cdot\mathbf{N} = \mathbf{V}\cdot\left[\mathbf{V}^{-1}\mathbf{G}(\mathbf{V}^T)^{-1}\right]^{1/2}, \qquad (A3)$$

where **N** is chosen such that $\mathbf{L} \cdot \mathbf{L}^T = \mathbf{G}$ [62]. Using information of Table 1A, the calculated **V** and **L** matrices are for σ-type stretching modes are:

$$\mathbf{V}(\tilde{X}^2\Sigma^+, \tilde{A}^2\Pi, \tilde{B}^2\Sigma^+) \approx \begin{vmatrix} & Q_1 & Q_3 \\ \Delta r_{O-H} & 0.9982 & -0.0597, -0.0598, 0.0598 \\ \Delta r_{Sr-O} & 0.0087, 0.0095, 0.0093 & 0.99799 \end{vmatrix} \quad (A4)$$

$$\mathbf{L}(\tilde{X}^2\Sigma^+, \tilde{A}^2\Pi, \tilde{B}^2\Sigma^+) \approx \begin{vmatrix} & Q_1 & Q_3 \\ \Delta r_{O-H} & 1.0270 & 0.0023, 0.0025, 0.0025 \\ \Delta r_{Sr-O} & -0.0615 & 0.2648 \end{vmatrix} . \quad (A5)$$

As expected **V** and **L** a nearly identical for the three states. Substitution of **L** and **B** into Eq. 7 produces a unit **J** rotation matrices, as expected because the normal coordinates $Q_1$ and $Q_3$ are parallel for the three states. Substitution of **L**, **B**, and the bond lengths of Table A1 into Eq. 8 gives:

$$\mathbf{D}(\tilde{X}^2\Sigma^+, \tilde{A}^2\Pi) \approx \begin{vmatrix} -0.00019 \\ 0.07549 \end{vmatrix} , \quad (A6)$$

and

$$\mathbf{D}(\tilde{X}^2\Sigma^+, \tilde{B}^2\Sigma^+) \approx \begin{vmatrix} -0.00110 \\ 0.05261 \end{vmatrix} . \quad (A7)$$

The FCFs given in Table 4 were obtained solving the analytical equations of Ref.[49] using the vibrational frequencies of Table A1 along with the predicted **J** and **D** matrices.

**Appendix B: 2D-DVR calculations of the bending Franck-Condon factors.**

For a DVR calculation, a basis is chosen based on a model Hamiltonian (in this case the Hamiltonian for a two-dimensional harmonic oscillator),

$$\hat{H}^{HO} = -\frac{\hbar^2}{2}\left[\frac{\partial}{\partial \theta_x}G_{\theta\theta}(\theta_e)\frac{\partial}{\partial \theta_x} + \frac{\partial}{\partial \theta_y}G_{\theta\theta}(\theta_e)\frac{\partial}{\partial \theta_y}\right] + \frac{k}{2}(\theta_x^2 + \theta_y^2) \quad (B1)$$

which is referred to as the Finite Basis Representation (FBR). Rather than evaluating the Hamiltonian matrix in the FBR, the basis set is rotated to a representation in which the matrix representation of the chosen coordinate, in this case $\theta^2$, is diagonal. The square roots of the eigenvalues of the matrix representation of $\theta^2$ in the FBR provide the DVR points. In this work, the DVR method is implemented in a three step approach, similar to that originally used by Harris et al.[54]. Specifically, the matrix representation of

$$\theta^2 = \theta_x^2 + \theta_y^2 \quad (B2)$$

is constructed in the basis set consisting of the eigenstates of the two-dimensional harmonic oscillator, and the matrix is diagonalized (Step I). The square roots of the resulting eigenvalues are the grid points for the DVR and the matrix of eigenvectors, **U**, provides the transformation between the FBR and the DVR.[57] Next, a cubic spline interpolation of the DFT calculated electronic energies (Supplemental Material) is used to obtain the values of the potential energy at the grid points (Step II). Finally, (Step III) the matrix representation of

$$\hat{H}^{bend} = \hat{H}^{HO} - \frac{\hbar^2}{2}\frac{\partial}{\partial \theta_x}[G_{\theta\theta}(\theta) - G_{\theta\theta}(\theta_e)]\frac{\partial}{\partial \theta_x} + \frac{\partial}{\partial \theta_y}[G_{\theta\theta}(\theta) - G_{\theta\theta}(\theta_e)]\frac{\partial}{\partial \theta_y} + V(\theta_x, \theta_y) -$$

$$\frac{k}{2}(\theta_x^2 + \theta_y^2) \quad (B3)$$

is evaluated in the DVR. The corrections to the kinetic energy operator that appear in Eq. B3 come from the dependence of the G-matrix elements on the bending normal coordinate, which have been obtained analytically[59]. The Hamiltonain matrix is diagonalized to obtain the desired bend energies and wave functions.

In the DVR, the matrix elements of the potential are $V_{ij} = V(\theta_i)\delta_{i,j}$, which are the interpolated electronic energies, evaluated at the desired values of θ. To determine the required $\theta_i$ values, we first define a pair of dimensionless coordinates $Q_x$ and $Q_y$, where

$$Q_{x/y} = \sqrt{\frac{\omega_e}{\hbar G^e_{\theta\theta}}}(\pi - \theta_{x/y}) = \sqrt{\alpha}(\pi - \theta_{x/y}) \tag{B4}$$

$G^e_{\theta\theta}$ is the value of the G-matrix element for the bend, evaluated at $\theta_e = \pi$, and $\omega_e = \sqrt{kG^e_{\theta\theta}}$. These coordinates are replaced by the polar coordinates[63, 64] (i.e. $Q_x = Q\cos\phi$, $Q_y = Q\sin\phi$, $Q^2 = Q_x^2 + Q_y^2$, $\tan\phi = Q_y/Q_x$, and $Q^\pm = Q_x \pm iQ_y = Qe^{\pm i\phi}$). As such, $Q^2 = Q^+Q^-$, while

$$-\hbar^2\left(\frac{\partial^2}{\partial\theta_x^2} + \frac{\partial^2}{\partial\theta_y^2}\right) = \frac{1}{\alpha}(P_x^2 + P_y^2) = \frac{P^+P^-}{\alpha} \ . \tag{B5}$$

With this change of variables,

$$\hat{H}^{HO} = \frac{\hbar\omega_e}{2}[P^+P^- + Q^+Q^-] \tag{B6}$$

and the energy eigenvalues are expressed in terms of the vibrational quantum number, v, and a vibrational angular momentum quantum number, l:

$$\hat{H}^{HO}|v,l\rangle = \hbar\omega_e(v+1)|v,l\rangle . \tag{B7}$$

With these definitions in place, we need to diagonalize a matrix representation of $(\pi - \theta)^2$ in the basis of eigenstates of $\hat{H}^{HO}$, $|v,l\rangle$, for a chosen value of l. Because the potential is independent of ϕ, l is a good quantum number. Noting that $(\pi - \theta)^2 = \alpha Q^+Q^-$, the matrix elements for Step I are[63]:

$$\theta^2_{i,j} = \tfrac{1}{\alpha}\langle i,l|Q^+Q^-|j,l\rangle = \tfrac{\hbar^2}{\alpha}\left[(i+1)\delta_{ij} + \tfrac{1}{2}\sqrt{(i+l+2)(i-l+2)}\delta_{i,j+2} + \tfrac{1}{2}\sqrt{(i+l)(i-l)}\delta_{i,j-2}\right].$$
(B8)

The eigenvalues of this matrix provide the values of θ at which the potential is evaluated, while the eigenvectors, **U**, provide the transformation matrix between the FBR and the DVR, which are used to transform the matrix representation of the kinetic energy operator to the DVR. In this calculation, $l = 0$ or 1, and the matrix representation of $\theta^2$ is set up for $i$ and $j$ each ranging from 0 to 400, although only the states that correspond to θ < 90° are used in the calculation.

Setting up a matrix representation of $\hat{H}^{bend}$ requires the evaluation of the matrix representation of $\hat{P}^2$ in the DVR using[63]

$$P^2_{i,j} = \hbar^2\alpha\langle i,l|P^+P^-|j,l\rangle = \hbar^2\alpha\left[(i+1)\delta_{ij} - \tfrac{1}{2}\sqrt{(i+l+2)(i-l+2)}\delta_{i,j+2} - \tfrac{1}{2}\sqrt{(i+l)(i-l)}\delta_{i,j-2}\right].$$
(B9)

The matrix representation of $\hat{P}^2$ is then transformed into the coordinate representation using the DVR-FBR transformation matrix ( $\boldsymbol{P}^2_{DVR} = \boldsymbol{U}^{-1}\boldsymbol{P}^2_{FBR}\boldsymbol{U}$ ). Before constructing the Hamiltonian, it should be noted that its evaluation is complicated by the fact that $\hat{P}^2$ does not commute with the coordinate-dependent G-matrix element. This is addressed by taking advantage of the relationship

$$\hat{T} = -\tfrac{\hbar^2}{2}\left[\tfrac{\partial}{\partial\theta_x}G_{\theta\theta}(\theta)\tfrac{\partial}{\partial\theta_x} + \tfrac{\partial}{\partial\theta_y}G_{\theta\theta}(\theta)\tfrac{\partial}{\partial\theta_y}\right] = -\tfrac{\hbar^2}{4}\left[\left(\tfrac{\partial^2}{\partial\theta_x^2} + \tfrac{\partial^2}{\partial\theta_y^2}\right)G_{\theta\theta}(\theta) + G_{\theta\theta}(\theta)\left(\tfrac{\partial^2}{\partial\theta_x^2} + \tfrac{\partial^2}{\partial\theta_y^2}\right) - \tfrac{\partial^2 G_{\theta\theta}}{\partial\theta^2}\right] = \tfrac{1}{4}\left[\hat{P}^2 G_{\theta\theta}(\theta) + G_{\theta\theta}(\theta)\hat{P}^2 + \hbar^2\tfrac{\partial^2 G_{\theta\theta}}{\partial\theta^2}\right].$$
(B10)

Since the $\frac{\partial^2 G_{\theta\theta}}{\partial \theta^2}$ term is expected to generate a small contribution to the energy, it has not been included in the present calculation.

Based on Eq. (B10), the matrix representation of $\hat{H}^{bend}$ in the DVR is

$$H_{i,j}^{bend} = \frac{1}{4}[G_{\theta\theta}(\theta_j) + G_{\theta\theta}(\theta_i)]P_{i,j}^2 + V(\theta_i)\delta_{i,j}, \qquad (B11)$$

where $\theta_i$ represents the square root of the $i^{th}$ eigenvalue of $\theta^2$. Once constructed, $\mathbf{H}^{bend}$ is diagonalized, and its eigenvalues are the energies of the states of interest, while the eigenvectors provide the amplitude of the wave function at the DVR points. With these, the bend FCFs reported in Table 4 were determined by squaring the inner product of these eigenvectors, $\mathbf{E}^{DVR}$. For example,

$$\left|\langle \psi_{bend}(\tilde{X}\,^2\Sigma^+) | \psi_{bend}(\tilde{B}\,^2\Sigma^+, \tilde{A}\,^2\Pi) \rangle\right|^2 = \left(\mathbf{E}^{DVR}(\tilde{X}\,^2\Sigma^+) \cdot \mathbf{E}^{DVR}(\tilde{B}\,^2\Sigma^+, \tilde{A}\,^2\Pi)\right)^2. \qquad (B14)$$

The FCF for the symmetry forbidden $\tilde{A}\,^2\Pi_{1/2}(0,0,0) \to \tilde{X}\,^2\Sigma^+(0,1^1,0)$ and $\tilde{B}\,^2\Sigma^+(0,0,0) \to \tilde{X}\,^2\Sigma^+(0,1^1,0)$ transitions were calculated to be zero, as expected.

The predicted bending vibrational frequencies from the DFT calculation (Supplemental Material), DVR, and the experimental data for the $\tilde{X}\,^2\Sigma^+$, $\tilde{A}\,^2\Pi$, and $\tilde{B}\,^2\Sigma^+$ states are given in Table B1.

Table B1. The observed and predicted relative energies[a].

|  | $\tilde{X}\,^2\Sigma^+$ | | $\tilde{A}\,^2\Pi$ | | $\tilde{B}\,^2\Sigma^+$ | |
|---|---|---|---|---|---|---|
|  | $(0,1^1,0)$ | $(0,2^0,0)$ | $(0,1^1,0)$ | $(0,2^0,0)$ | $(0,1^1,0)$ | $(0,2^0,0)$ |
| DFT[b] | 315 | - | 347 | - | 373 | |
| 2D-DVR | 322 | 638 | 308 | 614 | 358 | 699 |
| Exp.[c] | 364[c] | 703[c] | 378[d] | - | 401[c] | 771[c] |
|  |  |  |  |  |  |  |

a) Energies in wavenumber relative to the (0,0,0) level.
b) Predicted harmonic frequency $\omega_2$. See Supplemental Material. $\omega_1$(O-H stretch): $\tilde{X}\,^2\Sigma^+$ (= 3812.77cm$^{-1}$); $\tilde{A}\,^2\Pi$ (=3817.69 cm$^{-1}$); $\tilde{B}\,^2\Sigma^+$(= 3807.14cm$^{-1}$); $\omega_3$(Sr-O stretch): $\tilde{X}\,^2\Sigma^+$ (= 532.88 cm$^{-1}$); $\tilde{A}\,^2\Pi$ (= 538.20cm$^{-1}$); $\tilde{B}\,^2\Sigma^+$ (= 520.23 cm$^{-1}$)
c) Ref. 26

d) Ref.28

**Appendix C: Estimation of $b_{v',v''}$ from molecule loss during laser cooling**

The signal to noise of the recorded DF spectra was only sufficient to provide an upper limit for the very small $b_{v',v''}$ values of the $2_2^0, 3_2^0$ $\tilde{A}\,^2\Pi_r - \tilde{X}\,^2\Sigma^+$ and $2_2^0, 3_2^0$ $\tilde{B}\,^2\Sigma^+ - \tilde{X}\,^2\Sigma^+$ bands. Since both $\tilde{X}\,^2\Sigma^+(0,2^0,0)$ and $\tilde{X}\,^2\Sigma^+(0,0,2)$ states could be important loss channels during the prospective MOT trapping experiments, it is crucial to assess the accuracy of our predictions for corresponding $b_{v',v''}$ values given in Table 4. As a test we re-analyzed the data associated with the recent SrOH beam laser cooling measurement[19]. In that experiment the loss of molecules to excited dark vibrational states was determined and those measurements can be used to precisely extract experimental branching ratios and benchmark our calculations. For the details of the experimental configuration and SrOH photon scattering results we refer the reader to Refs. 19 and 35. Briefly, in the presence of rotationally and electronically closed transitions, the loss of the molecular signal as a function of the scattered photon number can be modeled as the Bernoulli sequence with probability to return to the ground vibrational state after scattering $N$ photons

$$P_{(remain)} = (1-p)^N. \tag{C1}$$

In equation C1, $p$ is the probability to decay to the dark vibrational states after a single spontaneous emission. Monitoring the molecular loss with the radiative pressure beam deflection using the $\tilde{A}\,^2\Pi_{1/2} - \tilde{X}\,^2\Sigma^+$ transition, the combined branching ratio to the excited vibrational states of the $\tilde{X}\,^2\Sigma^+(0,0,1)$ state was measured[13] to be $(3\pm1)\times10^{-3}$. In a laser cooling configuration, after scattering $220^{+110}_{-60}$ photons on the $0_0^0$ $\tilde{A}^2\Pi_{1/2} - \tilde{X}^2\Sigma^+$ transition with the presence of the $3_1^0$ $\tilde{A}^2\Pi_{1/2} - \tilde{X}^2\Sigma^+$ repumping laser, $(9\pm2)\%$ molecules decay to the $\tilde{X}(0,2^0,0)$ excited bending state and $(39 \pm 2)\%$ molecules decay to other vibrational states unaddressed by the two repumping lasers (i.e. $\tilde{X}(0,0,2)$ and higher vibrational levels). Since our calculations indicate that only decays to (0,0,1), (0,0,2), and $(0,2^2,0)$ have $b_{v',v''}$ values greater than $1\times10^{-4}$, we approximate the probability to decay to (0,0,2) state to be 39% with the appropriate uncertainty. Therefore, we determine the probability of decaying to the excited vibrational levels and the associated $b_{v',v''}$ value as

$$VBR = 1 - P_{(remain)}^{\frac{1}{N}}. \tag{C2}$$

Using $N=220$ and the corresponding $P_{(remain)} = 91\%$ for the bending mode $\tilde{X}(0,2^0,0)$ state and $P_{(remain)} = 61\%$ for the $\tilde{X}(0,0,2)$ state, we obtain estimates for $b_{v',v''}$ associated with the $2_2^0$ $\tilde{A}\,^2\Pi_r - \tilde{X}\,^2\Sigma^+$ fluorescence emission of $4_{-1}^{+2}\times10^{-4}$ and $b_{v',v''}$ associated with the $3_2^0$ $\tilde{A}\,^2\Pi_r - \tilde{X}\,^2\Sigma^+$ fluorescence emission of $2.2_{-0.7}^{+0.9}\times10^{-3}$ These values are consistent with our theoretical values given in Table 4. Furthermore the total branching ratio for the combined decay to the (0,0,2) and $(0,2^0,0)$ of $2.6\times10^{-3}$ is in agreement with the previous measurement[13] of to be $(3\pm1)\times10^{-3}$ based upon radiative force deflection as described above.

**Appendix D: Supplemental material**

Supplemental data for this article are available on the ScienceDirect(www.sciencedirect.com) . Supplemental data associated with this article can be found, in the online version, at htpp://???

**Figure Caption**

**Figure 1.** The dispersed fluorescence measurement of the $\tilde{A}^2\Pi_{1/2}(0,0,0)$ state obtained by cw-laser excitation of the $Q_1(9/2)$ ($\nu$ = 14453.497 cm$^{-1}$) line of the $0_0^0$ $\tilde{A}^2\Pi_{1/2}$ - $\tilde{X}^2\Sigma^+$ band. A background spectrum recorded with the dye laser blocked was subtracted to account for overlapping the Sr $^3P_1 \rightarrow$ $^1S$ emission ($\lambda_{air}$ = 689.26 nm). The expected locations of the emission to the $\tilde{X}^2\Sigma^+(0,1^1,0)$ and $\tilde{X}^2\Sigma^+(0,2^0,0)$ states are indicated.

**Figure 2.** The dispersed fluorescence spectrum of the $\tilde{A}^2\Pi_{3/2}(0,0,0)$ states resulting from excitation of the $R_2(7/2)$ ($\nu$ = 14806.947 cm$^{-1}$) line of the $0_0^0$ $\tilde{A}^2\Pi_{3/2}$ - $\tilde{X}^2\Sigma^+$ band. A background spectrum recorded with the dye laser blocked was subtracted to account for overlapping the Sr $^3P_1 \rightarrow$ $^1S$ emission ($\lambda_{air}$ = 689.26 nm). The expected locations of the emission to the $\tilde{X}^2\Sigma^+(0,1^1,0)$ and $\tilde{X}^2\Sigma^+(0,2^0,0)$ states are indicated.

**Figure 3.** The dispersed fluorescence spectrum resulting from excitation of the $R_1(11/2)$ ($\nu$ = 16080.181 cm$^{-1}$) line of the $0_0^0$ $\tilde{B}^2\Sigma^+ \leftarrow \tilde{X}^2\Sigma^+$ band and corresponding assignment. The feature near 690 nm is the $^3P_1 \rightarrow$ $^1S$ metastable emission ($\lambda_{air}$ = 689.26 nm) of atomic Sr strontium. Unlike the dispersed fluorescence spectra associated with the $\tilde{A}^2\Pi_{1/2}$ (Figure 1) and $\tilde{A}^2\Pi_{3/2}$ (Figure 2) levels, a background spectrum with the dye laser blocked was not subtracted because the Sr $^3P_1 \rightarrow$ $^1S$ emission ($\lambda_{air}$ = 689.26 nm) does not overlap with SrOH emission features. The expected locations of the emission to the $\tilde{X}^2\Sigma^+(0,1^1,0)$ and $\tilde{X}^2\Sigma^+(0,2^0,0)$ states are indicated.

**Figure 4.** The observed and predicted excitation spectra near the origin of the $0_0^0$ $\tilde{A}\,^2\Pi_{1/2} - \tilde{X}\,^2\Sigma^+$ band recorded field-free and in the presence of a 477 G magnetic field oriented perpendicular and parallel to the electric field of the laser radiation.

**Figure 5.** The observed and predicted excitation spectra near the origin of the $0_0^0$ $\tilde{A}\,^2\Pi_{3/2} - \tilde{X}\,^2\Sigma^+$ band recorded field-free and in the presence of a 477 G magnetic field orientated perpendicular and parallel to the electric field of the laser radiation.

**Figure 6.** The observed and predicted excitation spectra near the origin of the $0_0^0$ $\tilde{B}\,^2\Sigma^+ - \tilde{X}\,^2\Sigma^+$ band recorded field-free and in the presence of a 913 G magnetic field orientated perpendicular and parallel to the electric field of the laser radiation.

**Figure 7.** The predicted Zeeman tuning of the energy levels associated with low-$J$ branch features of the $0_0^0$ $\tilde{A}\,^2\Pi_{1/2} - \tilde{X}\,^2\Sigma^+$ band and associated assignment.

**Figure 8.** The predicted Zeeman tuning of the energy levels associated with low-$J$ branch features of the $0_0^0$ $\tilde{A}\,^2\Pi_{3/2} - \tilde{X}\,^2\Sigma^+$ band and associated assignment.

**Figure 9.** The predicted Zeeman tuning of the energy levels associated with low-$J$ branch features of the $0_0^0$ $\tilde{B}\,^2\Sigma^+ - \tilde{X}\,^2\Sigma^+$ band and associated assignment.

**Table 1.** Observed and calculated transition wavenumbers (cm$^{-1}$) for the $0_0^0$
$\tilde{A}\,^2\Pi_r \leftarrow \tilde{X}\,^2\Sigma^+$ band of SrOH.

| Branch[a] | Obs.[b] | Obs.-calc. | Branch | Obs. | Obs.-calc | Branch | Obs. | Obs.-calc |
|---|---|---|---|---|---|---|---|---|
| $P_2(2.5)$ | 803.6174 | 0.0008 | $R_2(0.5)$ | 806.1022 | -0.0004 | $P_1(1.5)$ | 542.3366 | 0.0005 |
| $P_2(3.5)$ | 802.8955 | 0.0010 | $R_2(1.5)$ | 806.3749 | 0.0002 | $P_1(2.5)$ | 542.1730 | 0.0001 |
| $P_2(4.5)$ | 802.1824 | 0.0004 | $R_2(2.5)$ | 806.6574 | 0.0009 | $P_1(3.5)$ | 542.0186 | 0.0001 |
|  |  |  | $R_2(3.5)$ | 806.9582 | -0.0007 | $P_1(4.5)$ | 541.8723 | -0.0003 |
| $^QP_{21}(2.5)$ | 805.1041 | -0.0004 | $R_2(4.5)$ | 807.2485 | -0.0006 | $P_1(5.5)$ | 541.7355 | 0.0000 |
| $^QP_{21}(3.5)$ | 804.8782 | -0.0001 | $R_2(5.5)$ | 807.5598 | -0.0001 | $P_1(6.5)$ | 541.6074 | 0.0004 |
| $^QP_{21}(4.5)$ | 804.6625 | 0.0007 | $R_2(6.5)$ | 807.8792 | -0.0011 | $P_1(7.5)$ | 541.4870 | -0.0002 |
| $^QP_{21}(5.5)$ | 804.4561 | 0.0012 |  |  |  | $P_1(8.5)$ | 541.3754 | -0.0006 |
| $^QP_{21}(6.5)$ | 804.2573 | -0.0005 | $Q_1(0.5)$ | 542.6884 | 0.0000 | $P_1(9.5)$ | 541.2735 | -0.0001 |
| $^QP_{21}(7.5)$ | 804.0693 | -0.0010 | $Q_1(1.5)$ | 542.8771 | -0.0004 | $P_1(10.5)$ | 541.1797 | -0.0001 |
| $^QP_{21}(10.5)$ | 803.5660 | -0.0001 | $Q_1(2.5)$ | 543.0753 | -0.0001 | $P_1(11.5)$ | 541.0946 | -0.0001 |
| $^QP_{21}(15.5)$ | 802.9205 | 0.0005 | $Q_1(3.5)$ | 543.2817 | -0.0001 | $P_1(12.5)$ | 541.0183 | 0.0001 |
|  |  |  | $Q_1(4.5)$ | 543.4968 | -0.0002 | $P_1(13.5)$ | 540.9509 | 0.0003 |
| $Q_2(1.5)$ | 805.1105 | -0.0001 | $Q_1(5.5)$ | 543.7204 | -0.0003 | $P_1(14.5)$ | 540.8911 | -0.0004 |
| $Q_2(2.5)$ | 804.8873 | 0.0005 | $Q_1(6.5)$ | 543.9531 | -0.0002 |  |  |  |
| $Q_2(3.5)$ | 804.6735 | 0.0008 | $Q_1(7.5)$ | 544.1945 | 0.0002 | $^PQ_{12}(0.5)$ | 542.3332 | 0.0007 |
| $Q_2(4.5)$ | 804.4686 | 0.0004 | $Q_1(8.5)$ | 544.4444 | 0.0003 | $^PQ_{12}(1.5)$ | 542.1670 | 0.0002 |
| $Q_2(5.5)$ | 804.2736 | 0.0001 |  |  |  | $^PQ_{12}(2.5)$ | 542.0098 | -0.0001 |
| $Q_2(6.5)$ | 804.0880 | -0.0005 | $^QR_{12}(0.5)$ | 542.8807 | -0.0005 | $^PQ_{12}(3.5)$ | 541.8621 | 0.0004 |
| $Q_2(9.5)$ | 803.5920 | 0.0005 | $^QR_{12}(1.5)$ | 543.0818 | 0.0004 | $^PQ_{12}(4.5)$ | 541.7221 | 0.0000 |
| $Q_2(14.5)$ | 802.9576 | 0.0001 | $^QR_{12}(2.5)$ | 543.2904 | 0.0000 | $^PQ_{12}(5.5)$ | 541.5917 | 0.0004 |
|  |  |  | $^QR_{12}(3.5)$ | 543.5079 | 0.0000 | $^PQ_{12}(6.5)$ | 541.4691 | 0.0001 |
| $^SR_{21}(0.5)$ | 806.6025 | 0.0003 | $^QR_{12}(4.5)$ | 543.7341 | 0.0000 | $^PQ_{12}(7.5)$ | 541.3553 | -0.0001 |
| $^SR_{21}(1.5)$ | 807.3724 | -0.0003 | $^QR_{12}(5.5)$ | 543.9692 | 0.0002 | $^PQ_{12}(8.5)$ | 541.2507 | 0.0002 |
| $^RQ_{21}(1.5)$ | 806.1060 | -0.0002 | $^QR_{12}(6.5)$ | 544.2123 | -0.0003 | $^PQ_{12}(9.5)$ | 541.1545 | 0.0001 |
| $^RQ_{21}(2.5)$ | 806.3802 | -0.0006 | $^QR_{12}(7.5)$ | 544.4649 | 0.0002 | $^PQ_{12}(10.5)$ | 541.0667 | -0.0002 |
| $^RQ_{21}(3.5)$ | 806.6656 | 0.0005 |  |  |  | $^PQ_{12}(11.5)$ | 540.9883 | 0.0002 |
| $^RQ_{21}(4.5)$ | 806.9471 | -0.0009 | $R_1(0.5)$ | 543.6644 | -0.0001 | $^PQ_{12}(12.5)$ | 540.9181 | 0.0002 |
| $^RQ_{21}(5.5)$ | 807.2615 | -0.0010 | $R_1(2.5)$ | 544.5042 | -0.0001 |  |  |  |
| $^RQ_{21}(6.5)$ | 807.5754 | -0.0002 |  |  |  | $^OP_{12}(1.5)$ | 541.1969 | 0.0000 |
|  |  |  |  |  |  | $^OP_{12}(2.5)$ | 540.3912 | -0.0004 |
| Std. dev. of fit = 0.00044 cm$^{-1}$ |

[a] A $^{\Delta N}\Delta J_{F_i'F_i''}(J'')$ branch designation is used. [b] The transition wavenumber – 14000cm$^{-1}$.

**Table 2.** Observed and calculated transition wavenumbers (cm$^{-1}$) for the $0_0^0$ $B$ $^2\Sigma^+$-$X^2\Sigma^+$ band system of SrOH.

| Branch[a] | Obs.[b] | Obs.-calc. | Branch[a] | Obs.[b] | Obs.-calc. |
|---|---|---|---|---|---|
| $R_1(1/2)$ | 77.9305 | -0.0009 | $^PQ_{12}(1/2)$ | 77.0028 | 0.0005 |
| $R_1(3/2)$ | 78.3689 | -0.0004 | $^PQ_{12}(5/2)$ | 76.4403 | 0.0004 |
| $R_1(5/2)$ | 78.8145 | 0.0013 | $^PQ_{12}(2.5)$ | 75.8827 | -0.0007 |
| $R_1(7/2)$ | 79.2627 | -0.0004 | $^PQ_{12}(7/2)$ | 75.3325 | -0.0005 |
| $R_1(9/2)$ | 79.7201 | 0.0011 | $^PQ_{12}(9/2)$ | 74.7900 | 0.0014 |
| $R_1(11/2)$ | 80.1815 | 0.0006 | $^PQ_{12}(11/2)$ | 74.2503 | 0.0000 |
| $R_1(13/2)$ | 80.6481 | -0.0007 | $^PQ_{12}(13/2)$ | 73.7171 | -0.0009 |
| $R_1(15/2)$ | 81.1234 | 0.0007 | | | |
| | | | | | |
| $R_2(1/2)$ | 78.7293 | -0.0001 | $P_1(3/2)$ | 76.9983 | -0.0003 |
| $R_2(3/2)$ | 79.3189 | 0.0005 | $P_1(5/2)$ | 76.4353 | 0.0016 |
| $R_2(5/2)$ | 79.9147 | 0.0014 | $P_1(7/2)$ | 75.8734 | -0.0015 |
| $R_2(7/2)$ | 80.5143 | 0.0001 | $P_1(9/2)$ | 75.3220 | -0.0001 |
| $R_2(9/2)$ | 81.1203 | -0.0008 | $P_1(11/2)$ | 74.7747 | -0.0005 |
| | | | $P_1(13/2)$ | 74.2342 | -0.0004 |
| | | | $P_1(15/2$ | 73.6996 | -0.0003 |
| $^RQ_{21}(1/2)$ | 78.1457 | 0.0004 | | | |
| $^RQ_{21}(3/2)$ | 78.7248 | -0.0009 | $P_2(3/2)$ | 76.6527 | -0.0011 |
| $^RQ_{21}(5/2)$ | 79.3129 | 0.0006 | $P_2(5/2)$ | 76.2403 | 0.0004 |
| $^RQ_{21}(7/2)$ | 79.9059 | 0.0011 | $P_2(7/2)$ | 75.8309 | -0.0011 |
| | | | $P_2(9/2)$ | 75.4305 | 0.0002 |
| | | | $P_2(11/2)$ | 75.0347 | 0.0002 |
| | | | $P_2(13/2)$ | 74.6445 | -0.0003 |
| | | | $P_2(15/2)$ | 74.2601 | -0.0010 |
| | | | $P_2(17/2)$ | 73.8831 | -0.0005 |
| | | | $P_2(19/2)$ | 73.5127 | 0.0007 |
| Std. dev. of fit = 0.00082 cm$^{-1}$ | | | | | |

[a] A $^{\Delta N}\Delta J_{F_i'F_i''}(J'')$ branch designation is used.
[b] The transition wavenumber -16300 cm$^{-1}$.

**Table 3.** Spectroscopic parameter's for the $\tilde{A}^2\Pi_r(0,0,0)$ and $\tilde{B}^2\Sigma^+(0,0,0)$ states of SrOH

| | $\tilde{A}^2\Pi_r(0,0,0)$ | | | $\tilde{B}^2\Sigma^+(0,0,0)$ | |
|---|---|---|---|---|---|
| Parameter[a] | This work [b,c] | Previous[d] | Parameter | This work [b,c] | Previous[e] |
| $A$ | 263.58741(20) | 263.58782(61) | $B$ | 0.2522066(89) | 0.25224(2) |
| $B$ | 0.2537833(16) | 0.25378477(36) | $\gamma$ | -0.142583(80) | -0.1447(3) |
| $A_D$ | $1.331\times10^{-5}$(fix) | $1.331\times10^{-5}$ | $D$ | $2.15\times10^{-7}$(fix) | $2.15\times10^{-7}$ |
| $D$ | $2.16859\times10^{-7}$(Fix) | $2.16859\times10^{-7}$ | $T_0(B\ ^2\Sigma^+)$ | 16377.49826(28) | 16377.505(1) |
| $p+2q$ | -0.143662(37) | -0.143595(14) | | | |
| $q$ | $-1.528\times10^{-4}$(fix) | $-1.528(19)\times10^{-4}$ | | | |
| $T_0(A\ ^2\Pi)$ | 14674.30016(10) | 14674.04171(39) | | | |
| Std. dev. of field-free fit =0.00046 cm$^{-1}$ | | | Std. dev. of field-free fit = 0.00090 cm$^{-1}$ | | |
| $g_l$ [f] | -0.269(22) | | | 0.306(18) | |
| Std. dev. of Zeeman fit = 28 MHz | | | Std. dev. of Zeeman fit = 31 MHz | | |

[a] Units for field-free parameters are wavenumbers (cm$^{-1}$).

[b] Numbers in parentheses represent a 2σ error estimate in the last quoted decimal point.

[c] The $X\ ^2\Sigma^+(v=0)$ state parameters were held fixed to those of Ref. 12: $B$= 0.249199814 cm$^{-1}$, $D$= $2.17437\times10^{-7}$ cm$^{-1}$, $\gamma$ =$2.42748\times10^{-3}$ cm$^{-1}$ $A_D$, $D$, and $q$ for the $\tilde{A}^2\Pi_r(0,0,0)$ state are fixed to those of Ref. 12 and for the $\tilde{B}^2\Sigma^+(0,0,0)$ state is fixed to that of Ref. 17.

[e] Ref. 17.

[f] $g_S$, and $g'_L$ constrained to 2.0023 and 1.000, respectively. $g_l$ for the $\tilde{X}\ ^2\Sigma^+(0,0,0)$ state constrained to -4.87x10$^{-3}$ (see text) and $g'$ to 0.00.

**Table 4.** Franck-Condon factors and branching ratios

| Band | $T_{v',v''}$ | Calc. FCFs | | Branching Ratios | |
|---|---|---|---|---|---|
| | | Stretch[a] | Bend[b] | Calc.[c] | Obs. |
| $0_0^0\ \tilde{A}^2\Pi_{1/2} - \tilde{X}^2\Sigma^+$ | 14546 | 0.9556 | 0.9995 | 0.9552 | 0.957±0.003 |
| $3_1^0\ \tilde{A}^2\Pi_{1/2} - \tilde{X}^2\Sigma^+$ | 14024 | 0.0425 | 0.9995 | 0.0425 | 0.043±0.002 |
| $3_2^0\ \tilde{A}^2\Pi_{1/2} - \tilde{X}^2\Sigma^+$ | 14182 | 0.0017 | 0.9995 | 0.0017 | < 0.005 |
| $1_1^0\ \tilde{A}^2\Pi_{1/2} - \tilde{X}^2\Sigma^+$ | (10768)[d] | $1.7\times10^{-6}$ | 0.9995 | $1.7\times10^{-6}$ | - |
| $2_2^0\ \tilde{A}^2\Pi_{1/2} - \tilde{X}^2\Sigma^+$ | 13818 | 0.9556 | 0.0005 | 0.0004 | < 0.005 |
| $0_0^0\ \tilde{A}^2\Pi_{3/2} - \tilde{X}^2\Sigma^+$ | 14805 | 0.9556 | 0.9995 | 0.9552 | 0.959±0.003 |
| $3_1^0\ \tilde{A}^2\Pi_{3/2} - \tilde{X}^2\Sigma^+$ | 14287 | 0.0425 | 0.9995 | 0.0425 | 0.041±0.004 |
| $3_2^0\ \tilde{A}^2\Pi_{3/2} - \tilde{X}^2\Sigma^+$ | 14411 | 0.0017 | 0.9995 | 0.0017 | < 0.007 |
| $1_1^0\ \tilde{A}^2\Pi_{3/2} - \tilde{X}^2\Sigma^+$ | (11027)[d] | $1.7\times10^{-6}$ | 0.9995 | $1.7\times10^{-6}$ | - |
| $2_2^0\ \tilde{A}^2\Pi_{3/2} - \tilde{X}^2\Sigma^+$ | 14017 | 0.9556 | 0.0004 | 0.0004 | < 0.007 |
| $0_0^0\ \tilde{B}^2\Sigma^+ - \tilde{X}^2\Sigma^+$ | 16380 | 0.9782 | 0.9973 | 0.9756 | 0.977±0.002 |
| $3_1^0\ \tilde{B}^2\Sigma^+ - \tilde{X}^2\Sigma^+$ | 15853 | 0.0212 | 0.9973 | 0.0211 | 0.023±0.003 |
| $3_2^0\ \tilde{B}^2\Sigma^+ - \tilde{X}^2\Sigma^+$ | 16016 | 0.0005 | 0.9973 | 0.0005 | < 0.005 |
| $1_1^0\ \tilde{B}^2\Sigma^+ - \tilde{X}^2\Sigma^+$ | (12602)[d] | $6.7\times10^{-5}$ | 0.9973 | $6.7\times10^{-5}$ | - |
| $2_2^0\ \tilde{B}^2\Sigma^+ - \tilde{X}^2\Sigma^+$ | 15653 | 0.9782 | 0.0027 | 0.0026 | < 0.005 |

a) $\left|\left\langle \tilde{A}^2\Pi, v_1, v_3 \middle| \tilde{X}^2\Sigma^+, v_1, v_3 \right\rangle\right|^2$ and $\left|\left\langle \tilde{B}^2\Sigma^+, v_1, v_3 \middle| \tilde{X}^2\Sigma^+, v_1, v_3 \right\rangle\right|^2$ values obtained using a two dimensional normal coordinate analysis (see text).

b) $\left|\left\langle \tilde{A}^2\Pi, v_2 \middle| \tilde{X}^2\Sigma^+, v_2 \right\rangle\right|^2$ and $\left|\left\langle \tilde{B}^2\Sigma^+, v_2 \middle| \tilde{X}^2\Sigma^+, v_2 \right\rangle\right|^2$ values obtained using a two-dimensional (2D) DVR method(see text).

c) The product of $\sigma$-symmetry stretching FCFs and bending FCFs.

d) Estimated using the 3778 cm$^{-1}$ determined $\omega_1$(O-H stretch) value for the $\tilde{X}^2\Sigma^+$ state of CaOH[44]. The ICCD is insensitive to emission associated with the $1_1^0$ bands.

# Figure 1

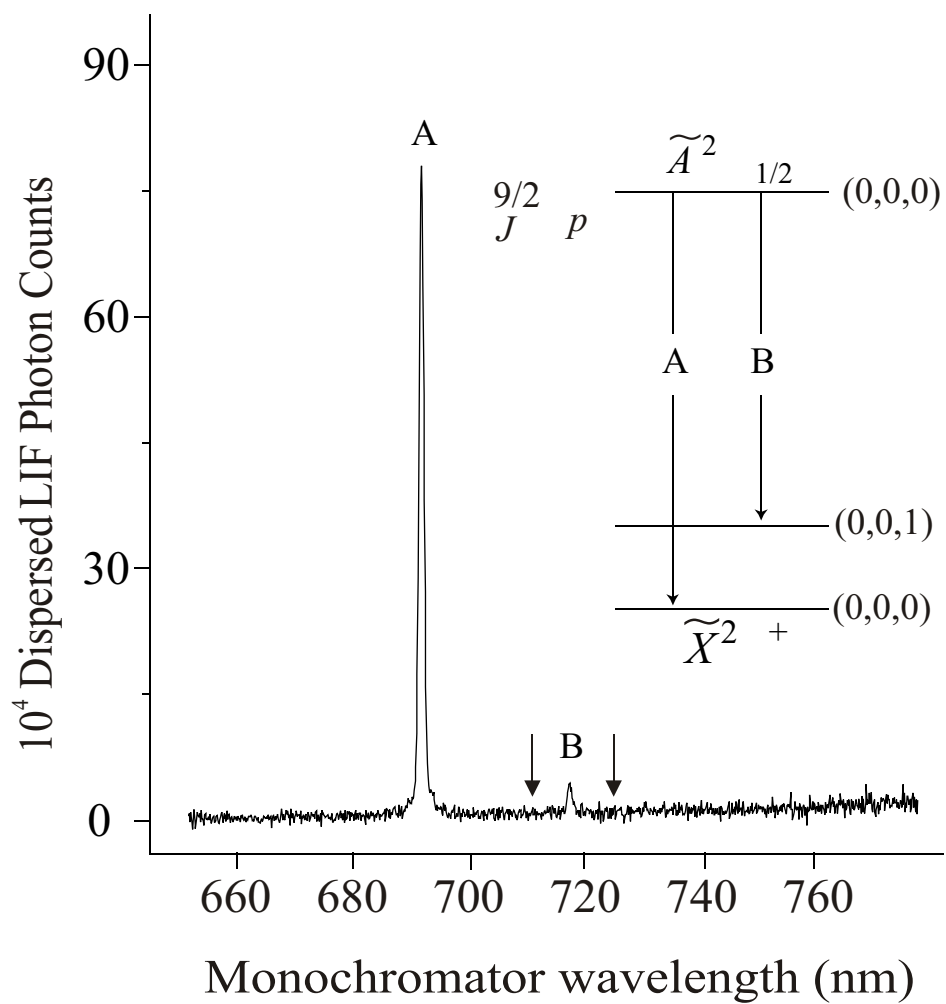

# Figure 2

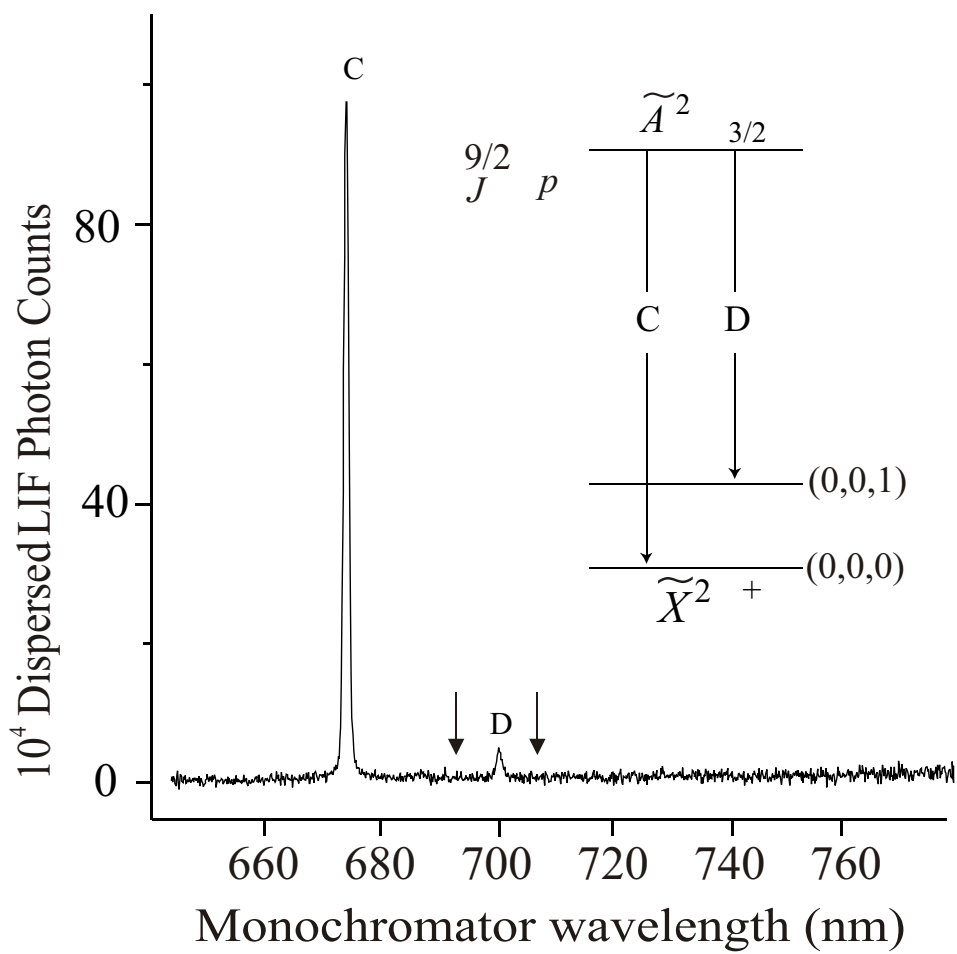

Figure 3

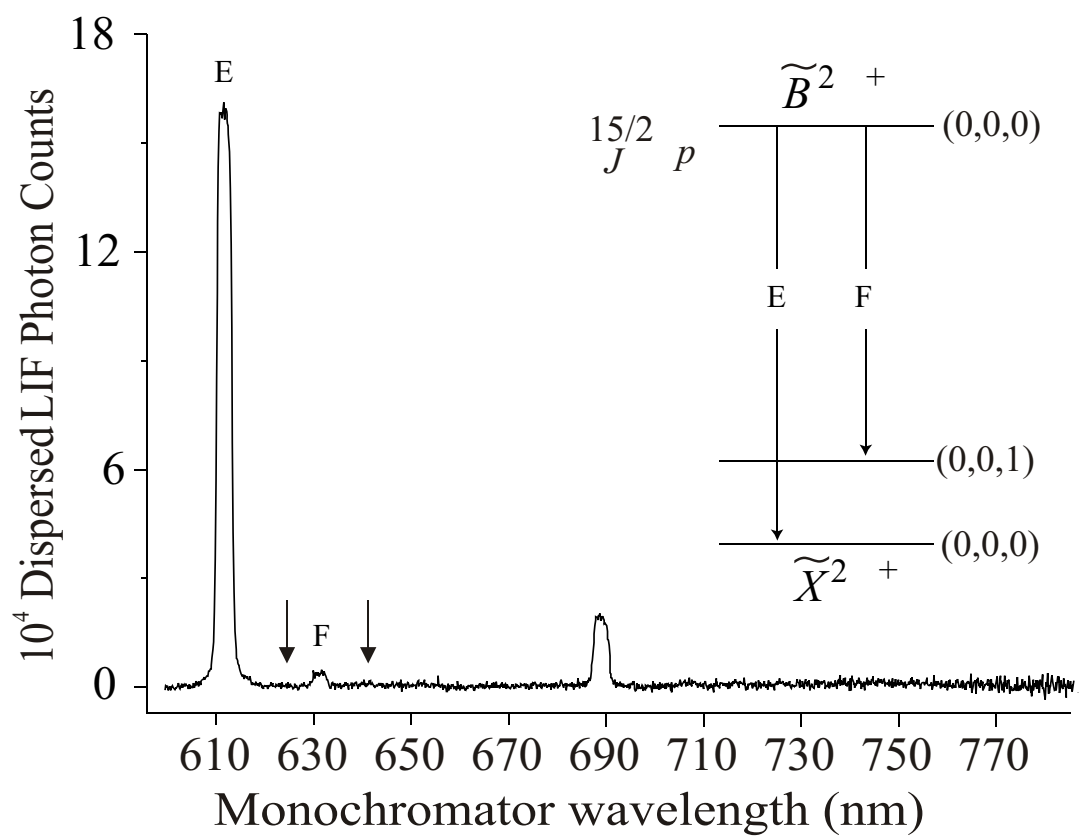

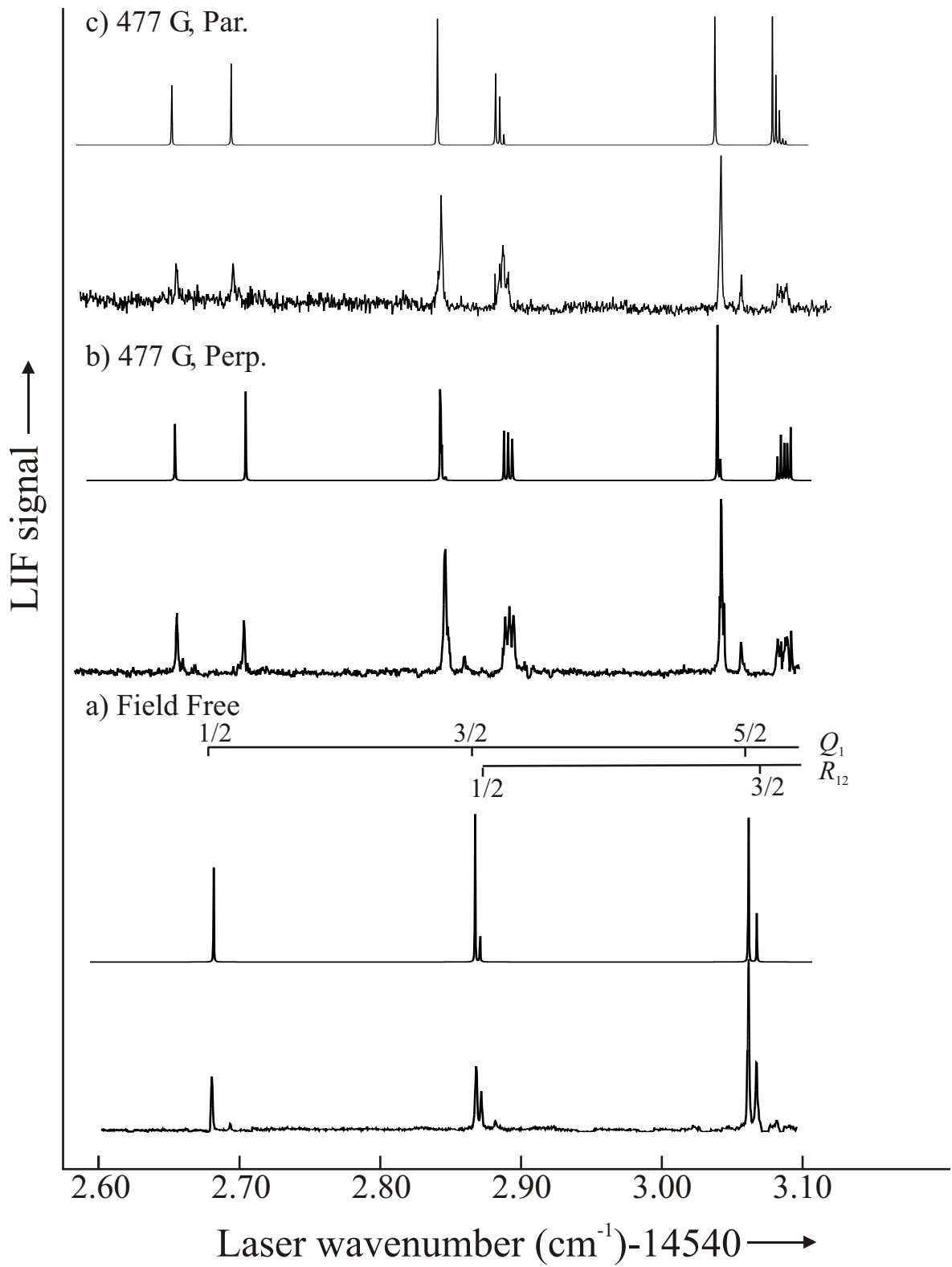

Figure 4

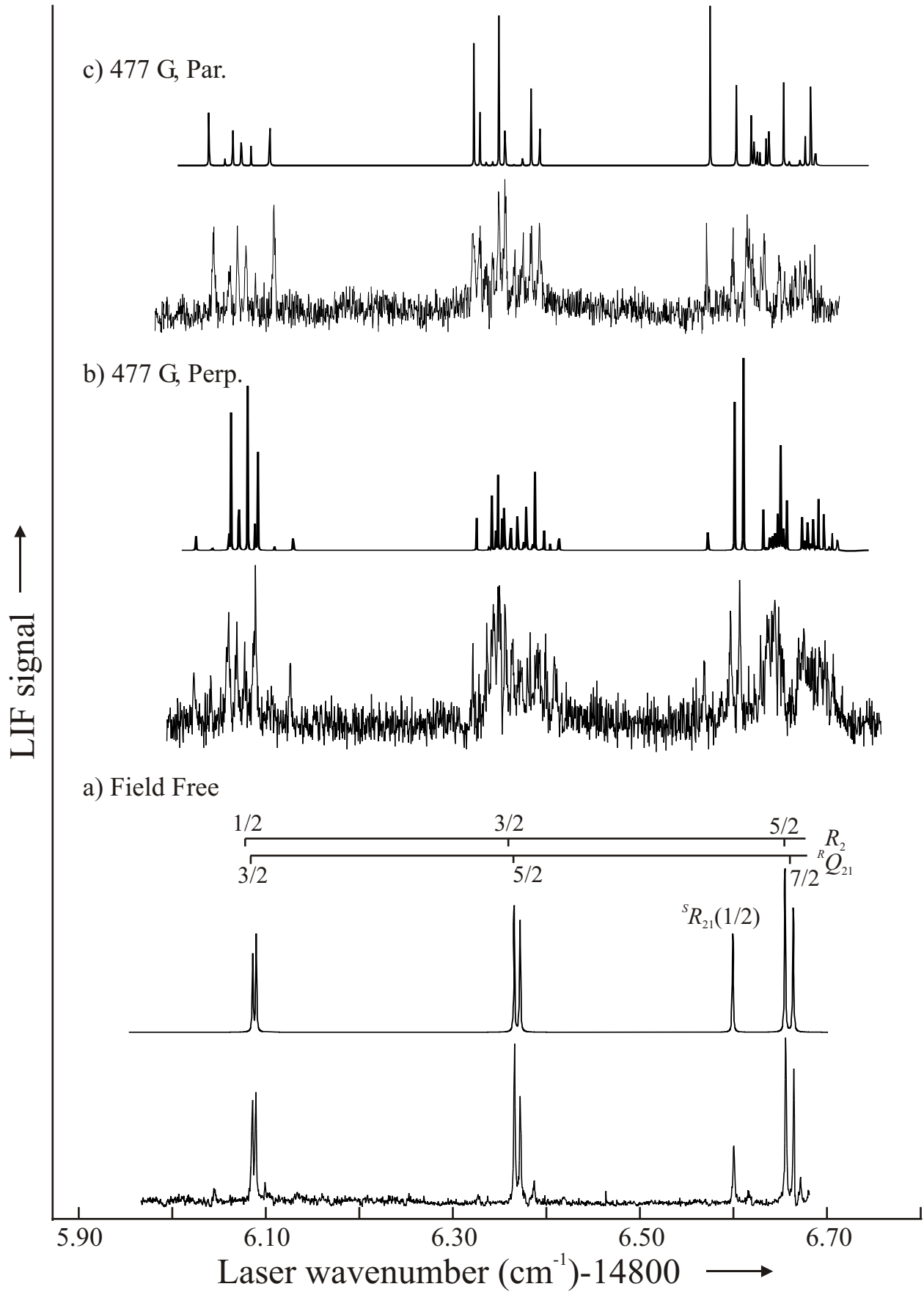

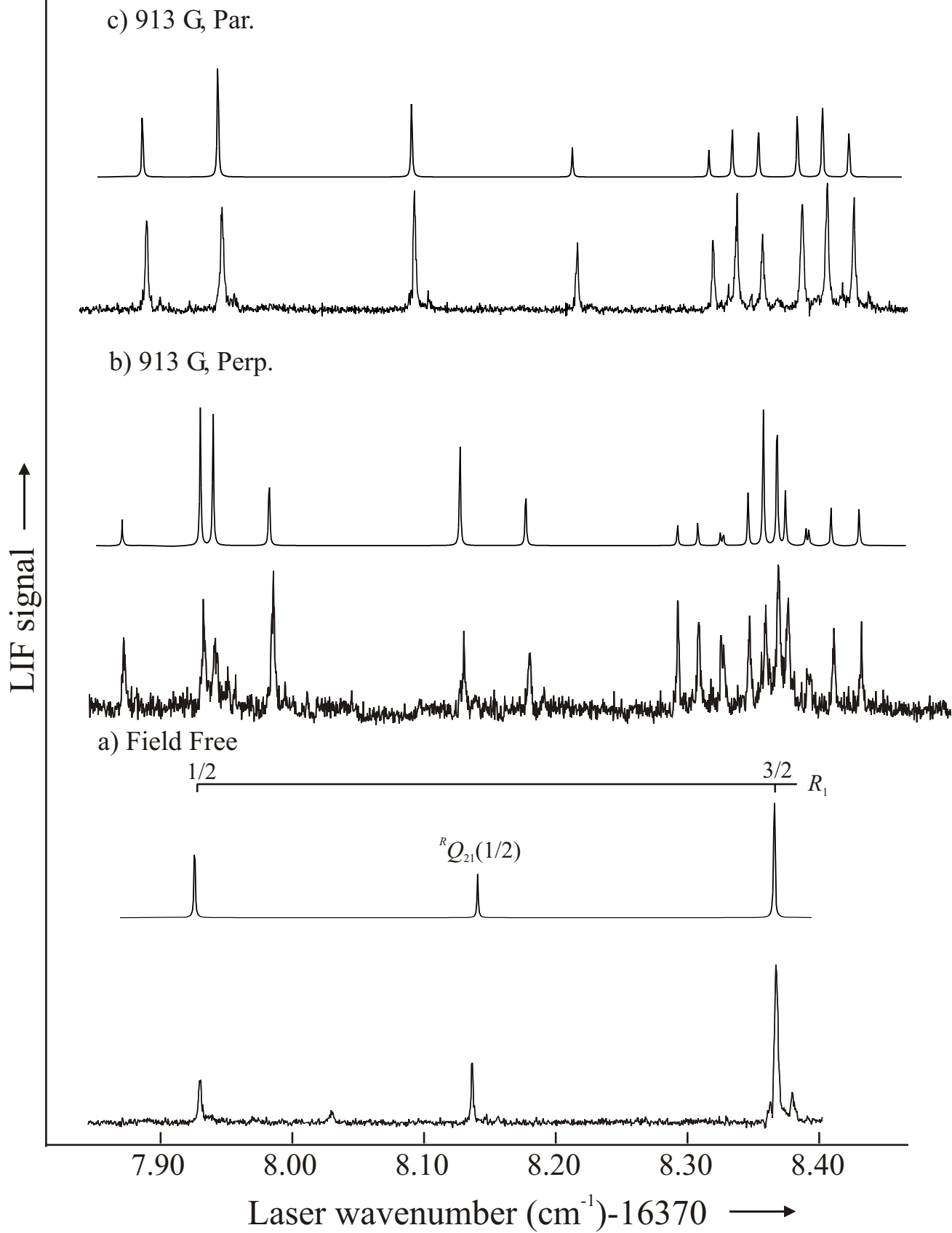

Figure 6

Figure 7

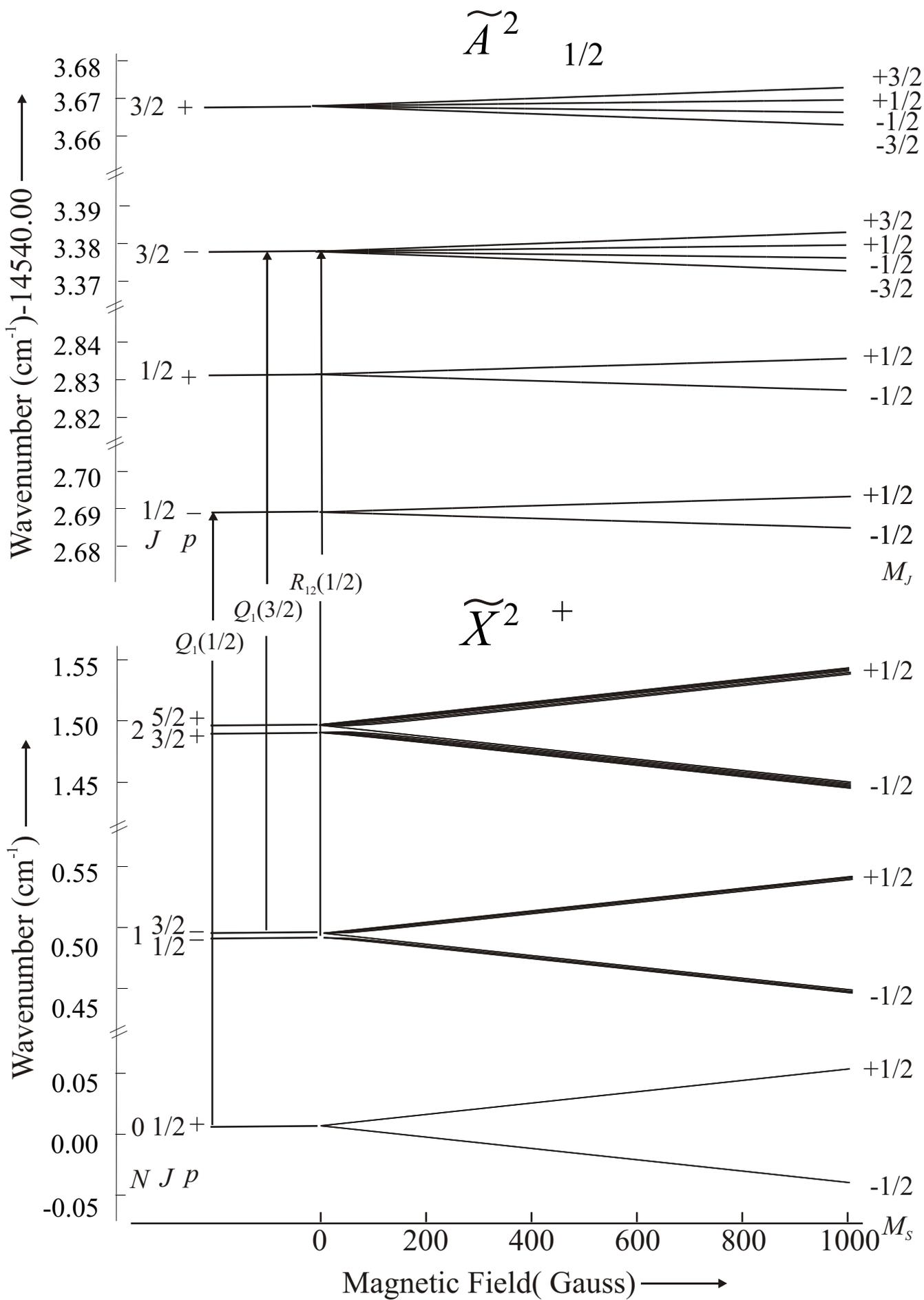

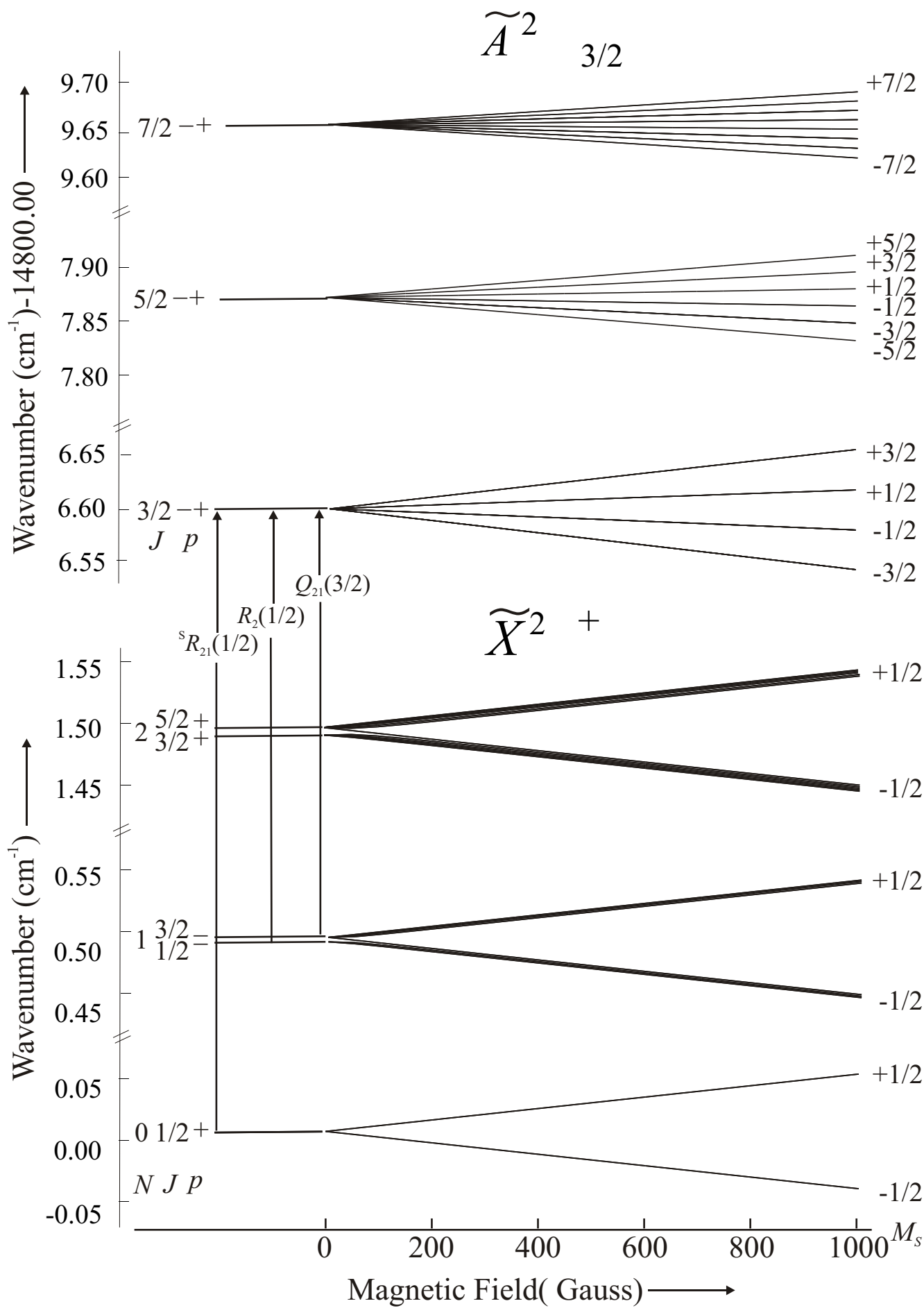

Figure 8

Figure 9

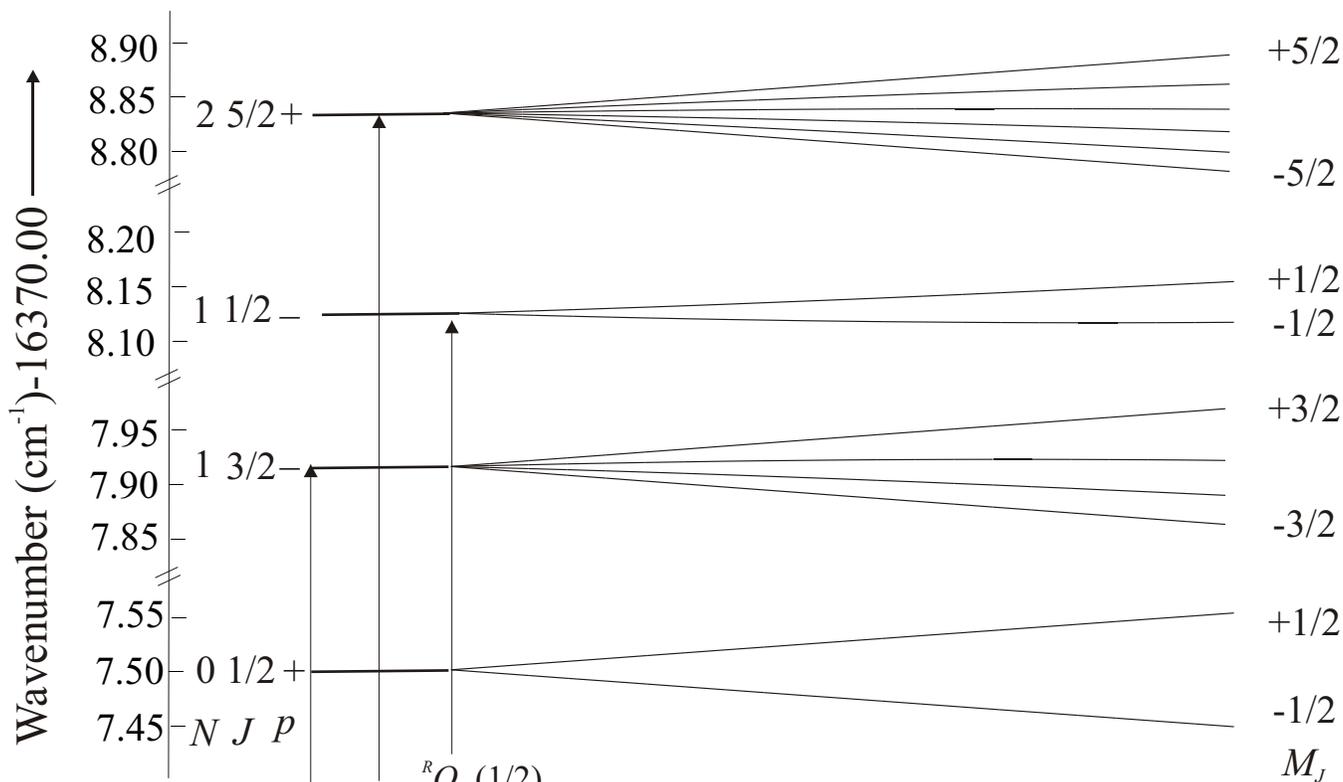
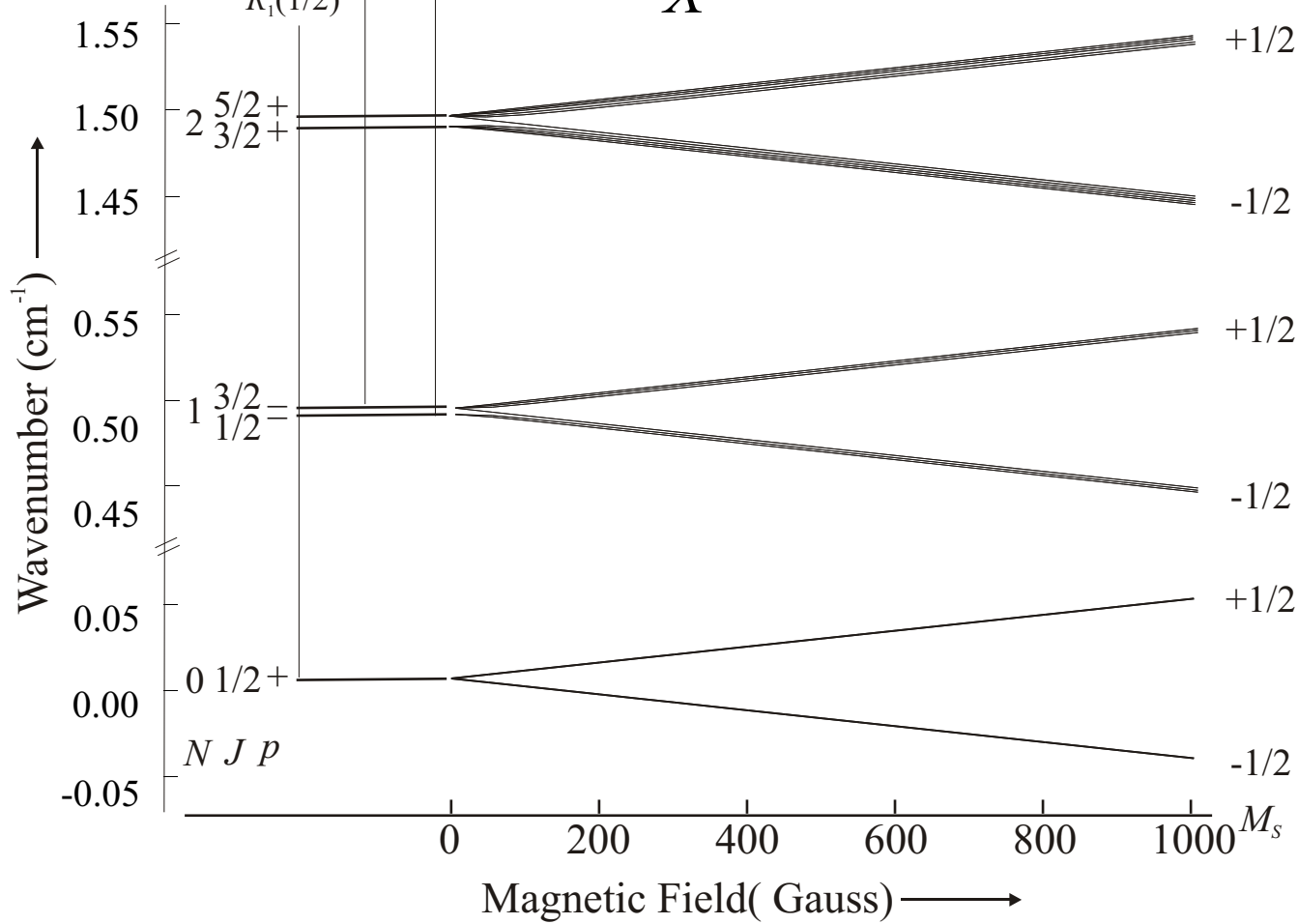

**Computational methods:**

All calculations were performed with the electronic structure program ORCA[1]. Briefly, the method consisted of the following steps:

i. The properties of the $\tilde{X}\,^2\Sigma^+$ state was predicted using DFT /BP86 functional.
ii. The properties of the $\tilde{A}\,^2\Pi$ and $\tilde{B}\,^2\Sigma^+$ states were predicted using TD-DFT . Only the first 10 excitations were allowed.
iii. To determine the potential energy as a function of angle, the angle of the $\tilde{X}\,^2\Sigma^+$ was varied in steps of 1° and the Sr-O and O-H bond lengths optimized. That optimized structure was then employed in the TD-DFT energy predictions of the $\tilde{A}\,^2\Pi$ and $\tilde{B}\,^2\Sigma^+$ states. The Sr-O and O-H bond lengths of the $\tilde{A}\,^2\Pi$ and $\tilde{B}\,^2\Sigma^+$ states were optimized with the angle constrained.

The numerical thresholds were systematically increased using the "TightOpt", "TightSCF", "Grid5" and "FinalGrid6" options. We carefully studied the accuracy and precise of our calculation by changing the basis set, functional that were all integrated in the ORCA package. In the end the def2-QZVPP[2], and corresponding def2-ECP along with the Weigend's "universal" Coulomb fitting (def2/J)[3] basis sets of the ORCA library were used. ECP parameters for Sr [Def2-ECP][4] have been obtained from TURBOMOLE (7.0.2)[5]. The predicted excitation energies of the $\tilde{A}\,^2\Pi$ and $\tilde{B}\,^2\Sigma^+$ states were 1.914 eV and 2.047 eV which compares favorably with the observed values of 1.819 eV and 2.030 eV. The predicted harmonic frequencies also compared favorably with observations (see Table below). The optimized energies and bond lengths as a function of bending angle are found in Table S4.

Predicted harmonic frequencies (cm$^{-1}$)

|  | $\tilde{X}\,^2\Sigma^+$ | $\tilde{A}\,^2\Pi$ | $\tilde{B}\,^2\Sigma^+$ |
|---|---|---|---|
| $\omega_1$(O-H stretch) | 3813 | 3818 | 3807 |
| $\omega_2$(Sr-O-H bend) | 315 | 347 | 373 |
| $\omega_3$(Sr-O stretch) | 533 | 538 | 520 |

References.

**Supplemental Table S1.** The observed and calculated Zeeman shifts of the $0_0^0$ $\tilde{A}\,^2\Pi_{1/2} - \tilde{X}\,^2\Sigma^+$ band.

| Branch, pol. | $\tilde{A}\,^2\Pi_{1/2}$ | | $\tilde{X}\,^2\Sigma^+$ | | | | Field-free (cm$^{-1}$) | Shift (MHz) | | Field |
|---|---|---|---|---|---|---|---|---|---|---|
| | J | M$_J$ | J | N | M$_S$ | M$_N$ | | Obs. | Obs.-Calc. | G |
| Q$_1$(1/2), ∥ | 1/2 | -1/2 | 1/2 | 0 | 1/2 | 0 | 14542.6884 | -1395 | -6 | 913 |
| | | 1/2 | 1/2 | | -1/2 | 0 | | 1421 | 33 | |
| Q$_1$(3/2), ∥ | 3/2 | -3/2 | 3/2 | 1 | 1/2 | -1 | 14542.8771 | -1333 | 12 | 913 |
| | | -1/2 | 3/2 | | 1/2 | 0 | | -1283 | 5 | |
| | | 1/2 | 3/2 | | 1/2 | 1 | | -1238 | -9 | |
| | | 3/2 | 3/2 | | 1/2 | 0 | | -1069 | 33 | |
| | | -1/2 | 3/2 | | -1/2 | 1 | | 1209 | -21 | |
| R$_{12}$(1/2), ⊥ | 3/2 | 1/2 | 1/2 | 1 | -1/2 | -1 | 14542.8806 | 1255 | 1 | 913 |
| | | 3/2 | 1/2 | | -1/2 | 0 | | 1379 | -2 | |
| Q$_1$(5/2) ∥ | 5/2 | -3/2 | 5/2 | 2 | 1/2 | -1 | 14543.0753 | -1241 | 22 | 913 |
| | | -3/2 | 5/2 | | -1/2 | 2 | | 1166 | -19 | |
| R$_{12}$(3/2), ⊥ | 5/2 | 1/2 | 3/2 | 2 | -1/2 | -1 | 14543.0818 | 1202 | -1 | 913 |
| | 5/2 | 3/2 | 3/2 | | -1/2 | 0 | | 1312 | 12 | |
| | 5/2 | 5/2 | 3/2 | | -1/2 | 1 | | 1415 | 20 | |
| R$_1$(1/2), ⊥ | 3/2 | -3/2 | 1/2 | 0 | 1/2 | 0 | 14543.0753 | -1415 | 0 | 913 |
| | 3/2 | 1/2 | 1/2 | | 1/2 | 0 | | -1222 | 9 | |
| | 3/2 | -1/2 | 1/2 | | -1/2 | 0 | | 1195 | -35 | |
| | 3/2 | 3/2 | 1/2 | | -1/2 | 0 | | 1400 | -14 | |
| Q$_1$(1/2), ∥ | 1/2 | -1/2 | 1/2 | 0 | 1/2 | 0 | 14542.6884 | -756 | -31 | 477 |
| | 1/2 | 1/2 | 1/2 | | -1/2 | 0 | | 718 | -7 | |
| | 3/2 | -3/2 | 3/2 | 1 | 1/2 | -1 | 14542.8771 | -654 | 16 | 477 |
| | 3/2 | 1/2 | 3/2 | | 1/2 | 1 | | -602 | 41 | |
| | 3/2 | -3/2 | 3/2 | | -1/2 | -1 | | 638 | -5 | |
| R$_{12}$(1/2), ⊥ | 3/2 | -1/2 | 1/2 | 1 | -1/2 | 1 | 14542.8806 | 487 | -38 | 477 |
| | 3/2 | 1/2 | 1/2 | | -1/2 | -1 | | 628 | 6 | |
| | 3/2 | 3/2 | 1/2 | | -1/2 | 0 | | 718 | 12 | |
| Q$_1$(3/2), ∥ | 3/2 | -3/2 | 3/2 | 1 | 1/2 | -1 | 14542.8771 | -654 | 16 | 477 |
| Q$_1$(5/2), ⊥ | 5/2 | 1/2 | 5/2 | 2 | 1/2 | 1 | 14543.0753 | -641 | -23 | 477 |
| | 5/2 | -1/2 | 5/2 | | 1/2 | 0 | | -641 | -25 | |
| | 5/2 | -3/2 | 5/2 | | 1/2 | -1 | | -641 | -29 | |
| | 5/2 | -3/2 | 5/2 | | -1/2 | 2 | | 587 | -32 | |
| R$_{12}$(3/2), ⊥ | 5/2 | -1/2 | 3/2 | 2 | -1/2 | -2 | 14542.8771 | 487 | -24 | 477 |
| | | 1/2 | 3/2 | | -1/2 | -1 | 14543.0818 | 590 | 9 | |
| | | 3/2 | 3/2 | | -1/2 | 0 | | 615 | -34 | |
| | | 5/2 | 3/2 | | -1/2 | 1 | | 705 | -9 | |
| Q$_1$(1/2), ∥ | 1/2 | 1/2 | 1/2 | 0 | 1/2 | 0 | 14542.6884 | -617 | -8 | 477 |
| | 1/2 | -1/2 | 1/2 | | -1/2 | 0 | | 592 | -16 | |
| Q$_1$(3/2) , ∥ | 3/2 | 3/2 | 3/2 | 1 | 1/2 | 1 | 14542.8771 | -617 | -23 | 477 |
| | 3/2 | 3/2 | 3/2 | | 1/2 | 1 | | -617 | -24 | |
| ∥ | 3/2 | -3/2 | 3/2 | | -1/2 | 1 | | 630 | 36 | |
| R$_{12}$(3/2) , ⊥∥ | 5/2 | -1/2 | 3/2 | 2 | -1/2 | -1 | 14543.0818 | 554 | -18 | 477 |
| | | 1/2 | 3/2 | | -1/2 | 0 | | 655 | -2 | |
| | | -3/2 | 3/2 | | -1/2 | -2 | | 533 | 55 | |
| | | 1/2 | 3/2 | | -1/2 | 0 | | 559 | -58 | |
| | | 3/2 | 3/2 | 2 | -1/2 | 1 | 14543.0818 | 635 | -47 | |

| $Q_1(5/2)$ , ‖ ‖ | 5/2 | 3/2 | 5/2 | 2 | 1/2 | 1 | 14543.0753 | -610 | -22 | 477 |
|---|---|---|---|---|---|---|---|---|---|---|
|  |  | 5/2 | 5/2 |  | 1/2 | 2 |  | -610 | -23 |  |
| ‖ |  | 5/2 | 5/2 |  | 1/2 | 2 |  | -610 | -24 |  |
|  |  | -5/2 | 5/2 |  | -1/2 | 2 |  | 597 | 10 |  |

**Supplemental Table S2.** The observed and calculated Zeeman shifts of the $0^0_0$ $\tilde{A}\,^2\Pi_{3/2} - \tilde{X}\,^2\Sigma^+$ band.

| Branch, pol. | $\tilde{A}\,^2\Pi_{3/2}$ | | $\tilde{X}\,^2\Sigma^+$ | | | | Field-free (cm$^{-1}$) | Shift (MHz) | | Field |
|---|---|---|---|---|---|---|---|---|---|---|
| | J | $M_J$ | J" | N | $M_S$ | N | | Obs. | Obs.-Calc. | G |
| $R_2(1/2)$ and $^RQ_{21}(3/2)$, ∥ | 3/2 | 1/2 | 3/2 | 1 | 1/2 | -1 | 14806.1022 | -910 | -41 | 477 |
| | | 1/2 | 3/2 | 1 | 1/2 | 0 | | -403 | -27 | |
| | | -3/2 | 3/2 | 1 | -1/2 | -1 | | -143 | -8 | |
| | | 3/2 | 3/2 | 1 | 1/2 | +1 | | 117 | -9 | |
| | | -1/2 | 1/2 | 1 | -1/2 | +1 | 14806.1060 | 325 | 2 | 477 |
| | | 1/2 | 1/2 | 1 | -1/2 | 0 | | 910 | 20 | |
| $R_2(3/2)$ and $^RQ_{21}(5/2)$, ∥ | 5/2 | -3/2 | 5/2 | 2 | 1/2 | -2 | 14806.3749 | -928 | -61 | 477 |
| | | -1/2 | 5/2 | 2 | 1/2 | -1 | | -724 | -47 | |
| | | 1/2 | 5/2 | 2 | 1/2 | 0 | | -521 | -34 | |
| | | 3/2 | 5/2 | 2 | 1/2 | 1 | | -317 | -22 | |
| | | 5/2 | 5/2 | 2 | 1/2 | 2 | | -114 | -12 | |
| | | -5/2 | 5/2 | 2 | -1/2 | -2 | | 89 | -7 | |
| | | -3/2 | 3/2 | 2 | -1/2 | 2 | 14806.3802 | 216 | 32 | 477 |
| | | -1/2 | 3/2 | 2 | -1/2 | 1 | | 483 | 32 | |
| | | 1/2 | 3/2 | 2 | -1/2 | 0 | | 737 | 22 | |
| | | 3/2 | 3/2 | 2 | -1/2 | -1 | | 991 | 16 | |
| $^SR_{21}(1/2)$, mixed | 3/2 | 1/2 | 1/2 | 0 | 1/2 | 0 | 14806.6025 | -422 | -13 | 477 |
| | | -1/2 | 1/2 | 0 | -1/2 | 0 | | 409 | 15 | |
| $R_2(5/2)$ and $^RQ_{21}(7/2)$, ∥ | 7/2 | -5/2 | 7/2 | 3 | 1/2 | -3 | 14806.6574 | -813 | -43 | 477 |
| | | -3/2 | 7/2 | 3 | 1/2 | -2 | | -736 | -52 | |
| | | -1/2 | 7/2 | 3 | 1/2 | -1 | | -654 | -49 | |
| | | 5/2 | 7/2 | 3 | 1/2 | 3 | | -349 | 4 | |
| | | 7/2 | 7/2 | 3 | 1/2 | 3 | | -258 | -28 | |
| | | -7/2 | 7/2 | 3 | -1/2 | 3 | | 181 | -44 | |
| | | -5/2 | 5/2 | 3 | -1/2 | -3 | 14806.6656 | 181 | 41 | 477 |
| | | -1/2 | 5/2 | 3 | -1/2 | -1 | | 530 | 59 | |
| | | 1/2 | 5/2 | 3 | -1/2 | 0 | | 685 | 51 | |
| | | 3/2 | 5/2 | 3 | -1/2 | 1 | | 840 | 47 | |
| | | 5/2 | 5/2 | 3 | -1/2 | 2 | | 987 | 38 | |
| $R_2(1/2)$ and $^RQ_{21}(3/2)$, ∥ | 3/2 | -1/2 | 3/2 | 1 | 1/2 | 0 | 14806.1022 | -884 | 24 | 477 |
| | 3/2 | 1/2 | 3/2 | 1 | 1/2 | -1 | | -364 | -25 | |
| | 3/2 | -3/2 | 3/2 | 1 | -1/2 | 0 | | -198 | 8 | |
| | 3/2 | 3/2 | 1/2 | 1 | 1/2 | 1 | | 143 | -15 | |
| | 3/2 | -1/2 | 1/2 | 1 | -1/2 | 0 | | 348 | -12 | |
| | 3/2 | 3/2 | 1/2 | 1 | -1/2 | 0 | | 1401 | -25 | |
| | 5/2 | -5/2 | 5/2 | 2 | 1/2 | -2 | 14806.1060 | -1079 | 17 | 477 |
| | 5/2 | 5/2 | 3/2 | 2 | -1/2 | -1 | | 1133 | -66 | |
| $^SR_{21}(1/2)$, ⊥ | 3/2 | -1/2 | 1/2 | 0 | 1/2 | 0 | 14806.6021 | -923 | 17 | 477 |
| | 3/2 | -3/2 | 1/2 | 0 | -1/2 | 0 | | -140 | -4 | |
| | 3/2 | 3/2 | 1/2 | 0 | -1/2 | 0 | | 120 | -5 | |
| $R_2(5/2)$ and $^RQ_{21}(7/2)$, ∥ | 7/2 | -7/2 | 7/2 | 3 | 1/2 | -3 | 14806.6656 | -897 | 0 | 477 |
| $R_2(7/2)$ and $^RQ_{21}(9/2)$, ∥ | 9/2 | 3/2 | 9/2 | 4 | -1/2 | 3 | 14806.9582 | 1050 | 59 | 477 |

**Supplemental Table S3.** The observed and calculated Zeeman shifts of the $0_0^0$ $\tilde{B}\,^2\Sigma^+ - \tilde{X}\,^2\Sigma^+$ band.

| Branch, pol | $\tilde{B}\,^2\Sigma^+$ | | | $\tilde{X}\,^2\Sigma^+$ | | | | Field-free (cm$^{-1}$) | Shift (MHz) | | Field G |
|---|---|---|---|---|---|---|---|---|---|---|---|
| | J | N | $M_{J'}$ | J | N | $M_S$ | $M_N$ | | Obs. | Obs.-Calc. | |
| $R_1(3/2), \perp$ | 5/2 | 2 | -3/2 | 3/2 | 1 | +1/2 | -1 | 16378.3689 | -2220 | -49 | 913 |
| | | | | -1/2 | 3/2 | | +1/2 | 1 | | -1705 | -13 | |
| | | | | 1/2 | 3/2 | | +1/2 | 1 | | -1182 | -17 | |
| | | | | 1/2 | 3/2 | | +1/2 | -1 | | -1107 | -13 | |
| | | | | 3/2 | 3/2 | | +1/2 | 0 | | -509 | 1 | |
| | | | | -5/2 | 1/2 | | -1/2 | +1 | | -170 | 0 | |
| | | | | 5/2 | 3/2 | | +1/2 | +1 | | 170 | 0 | |
| | | | | -3/2 | 1/2 | | -1/2 | -1 | | 268 | 27 | |
| | | | | -1/2 | 1/2 | | -1/2 | 0 | | 837 | 44 | |
| | | | | -1/2 | 1/2 | | -1/2 | +1 | | 796 | -32 | |
| | | | | 1/2 | 1/2 | | -1/2 | -1 | | 1347 | 28 | |
| | | | | 3/2 | 1/2 | | -1/2 | 0 | | 2001 | 26 | |
| $R_1(3/2), \parallel$ | 5/2 | 2 | -1/2 | 3/2 | 1 | +1/2 | -1 | 16378.3689 | -1643 | 10 | 913 |
| | | | | 1/2 | 3/2 | | +1/2 | 0 | | -1100 | 30 | |
| | | | | 3/2 | 3/2 | | +1/2 | +1 | | -543 | -1 | |
| | | | | -3/2 | 1/2 | | -1/2 | +1 | | -326 | 14 | |
| | | | | -1/2 | 1/2 | | -1/2 | -1 | | 769 | 12 | |
| | | | | 1/2 | 1/2 | | -1/2 | 0 | | 1373 | 19 | |
| $P_1(5/2)$, $^PQ_{12}(3/2) \parallel$ | 3/2 | 1 | -1/2 | 5/2 | 2 | +1/2 | -1 | 16376.4353 | -1916 | -47 | 913 |
| | | | | 1/2 | 5/2 | | +1/2 | 0 | | -1035 | -9 | |
| | | | | -3/2 | 3/2 | | -1/2 | -2 | 16376.4403 | -281 | 21 | |
| | | | | 3/2 | 5/2 | | +1/2 | +1 | 16376.4353 | 152 | -41 | |
| | | | | -1/2 | 3/2 | | -1/2 | -1 | 16376.4403 | 458 | -15 | |
| | | | | 1/2 | 3/2 | | -1/2 | 0 | | 1413 | 23 | |
| $P_2(3/2), \parallel$ | 1/2 | 1 | +1/2 | 5/2 | 2 | +1/2 | 0 | 16376.6527 | -1458 | 11 | 913 |
| | | | | -1/2 | 5/2 | | +1/2 | -1 | | -409 | -15 | |
| | | | | 1/2 | 3/2 | | -1/2 | 0 | | 948 | 0 | |
| | | | | -1/2 | 3/2 | | -1/2 | -1 | | 1993 | 44 | |
| $R_1(1/2) \perp$ | 3/2 | 1 | -1/2 | 1/2 | 0 | +1/2 | 0 | 16377.9305 | -2026 | -51 | 913 |
| | | | | -3/2 | 1/2 | | -1/2 | 0 | | -169 | -10 | |
| | | | | 3/2 | 1/2 | | +1/2 | 0 | | 174 | 16 | |
| | | | | 1/2 | 1/2 | | -1/2 | 0 | | 1494 | 36 | |
| $R_1(1/2), \parallel$ | 3/2 | 1 | 1/2 | 1/2 | 0 | +1/2 | 0 | | -1091 | 5 | |
| | | | | -1/2 | 1/2 | | -1/2 | 0 | | 577 | -1 | |
| $^QR_{12}(1/2), \parallel$ | 1/2 | 1 | 1/2 | 1/2 | 0 | 1/2 | 0 | 16378.1457 | -511 | -11 | 913 |
| | | | | -1/2 | 1/2 | | -1/2 | 0 | | 1058 | 41 | |
| $^QR_{12}(1/2), \perp$ | 1/2 | 1 | -1/2 | 1/2 | 0 | 1/2 | 0 | | -1523 | 14 | 913 |
| | | | | 1/2 | 1/2 | | -1/2 | 0 | | 2071 | 18 | |

Table S4. The DFT predicted energies and bond lengths as a function of bending angle.

| | $X^2\Sigma^+$ | | | | $A^2\Pi_r$ | | | | $B^2\Sigma^+$ | | |
|---|---|---|---|---|---|---|---|---|---|---|---|
| Angle (Degrees) | Energy (cm-1) | rSr-O (A°) | rOH (A°) | Angle (Degrees) | Energy (cm-1) | rSr-O (A°) | rOH (A°) | Angle (Degrees) | Energy (cm-1) | rSr-O (A°) | rOH (A°) |
| 90.00 | 3375.573 | 2.18 | 0.98 | 90.00 | 18497.234 | 2.19 | 0.98 | 90.00 | 20287.879 | 2.20 | 0.98 |
| 91.00 | 3273.129 | 2.18 | 0.98 | 91.00 | 18401.998 | 2.19 | 0.98 | 91.00 | 20178.602 | 2.20 | 0.98 |
| 92.00 | 3174.339 | 2.18 | 0.98 | 92.00 | 18310.096 | 2.19 | 0.98 | 92.00 | 20073.625 | 2.19 | 0.98 |
| 93.00 | 3079.052 | 2.18 | 0.98 | 93.00 | 18221.536 | 2.19 | 0.98 | 93.00 | 19972.465 | 2.19 | 0.98 |
| 94.00 | 2987.151 | 2.18 | 0.98 | 94.00 | 18136.341 | 2.19 | 0.98 | 94.00 | 19874.844 | 2.19 | 0.98 |
| 95.00 | 2898.501 | 2.18 | 0.98 | 95.00 | 18054.568 | 2.19 | 0.98 | 95.00 | 19781.422 | 2.19 | 0.98 |
| 96.00 | 2812.899 | 2.18 | 0.98 | 96.00 | 17976.067 | 2.18 | 0.98 | 96.00 | 19691.391 | 2.19 | 0.98 |
| 97.00 | 2730.084 | 2.17 | 0.98 | 97.00 | 17900.510 | 2.18 | 0.98 | 97.00 | 19604.269 | 2.18 | 0.98 |
| 98.00 | 2649.803 | 2.17 | 0.98 | 98.00 | 17827.279 | 2.18 | 0.98 | 98.00 | 19519.631 | 2.18 | 0.98 |
| 99.00 | 2571.900 | 2.17 | 0.98 | 99.00 | 17755.840 | 2.18 | 0.98 | 99.00 | 19437.270 | 2.18 | 0.98 |
| 100.00 | 2496.339 | 2.17 | 0.98 | 100.00 | 17685.851 | 2.18 | 0.98 | 100.00 | 19357.035 | 2.18 | 0.98 |
| 101.00 | 2423.161 | 2.17 | 0.98 | 101.00 | 17617.292 | 2.18 | 0.98 | 101.00 | 19278.396 | 2.18 | 0.97 |
| 102.00 | 2352.361 | 2.17 | 0.98 | 102.00 | 17550.268 | 2.17 | 0.98 | 102.00 | 19202.012 | 2.18 | 0.97 |
| 103.00 | 2283.846 | 2.17 | 0.97 | 103.00 | 17485.261 | 2.17 | 0.98 | 103.00 | 19127.627 | 2.18 | 0.97 |
| 104.00 | 2217.459 | 2.17 | 0.97 | 104.00 | 17422.591 | 2.17 | 0.98 | 104.00 | 19054.768 | 2.17 | 0.97 |
| 105.00 | 2152.987 | 2.17 | 0.97 | 105.00 | 17362.542 | 2.17 | 0.98 | 105.00 | 18983.691 | 2.17 | 0.97 |
| 106.00 | 2090.212 | 2.16 | 0.97 | 106.00 | 17305.303 | 2.17 | 0.98 | 106.00 | 18913.733 | 2.17 | 0.97 |
| 107.00 | 2028.996 | 2.16 | 0.97 | 107.00 | 17250.938 | 2.17 | 0.98 | 107.00 | 18845.687 | 2.17 | 0.97 |
| 108.00 | 1969.300 | 2.16 | 0.97 | 108.00 | 17199.131 | 2.16 | 0.98 | 108.00 | 18779.400 | 2.17 | 0.97 |
| 109.00 | 1911.163 | 2.16 | 0.97 | 109.00 | 17149.599 | 2.16 | 0.98 | 109.00 | 18714.829 | 2.17 | 0.97 |
| 110.00 | 1854.623 | 2.16 | 0.97 | 110.00 | 17101.775 | 2.16 | 0.98 | 110.00 | 18652.614 | 2.17 | 0.97 |
| 111.00 | 1799.631 | 2.16 | 0.97 | 111.00 | 17055.257 | 2.16 | 0.97 | 111.00 | 18592.029 | 2.16 | 0.97 |
| 112.00 | 1746.042 | 2.16 | 0.97 | 112.00 | 17009.449 | 2.16 | 0.97 | 112.00 | 18533.557 | 2.16 | 0.97 |
| 113.00 | 1693.663 | 2.16 | 0.97 | 113.00 | 16964.006 | 2.15 | 0.97 | 113.00 | 18476.323 | 2.16 | 0.97 |
| 114.00 | 1642.344 | 2.15 | 0.97 | 114.00 | 16918.690 | 2.15 | 0.97 | 114.00 | 18419.866 | 2.16 | 0.97 |
| 115.00 | 1592.027 | 2.15 | 0.97 | 115.00 | 16873.213 | 2.15 | 0.97 | 115.00 | 18364.292 | 2.16 | 0.97 |
| 116.00 | 1542.751 | 2.15 | 0.97 | 116.00 | 16828.169 | 2.15 | 0.97 | 116.00 | 18309.483 | 2.16 | 0.97 |
| 117.00 | 1494.608 | 2.15 | 0.97 | 117.00 | 16783.666 | 2.15 | 0.97 | 117.00 | 18255.544 | 2.15 | 0.97 |
| 118.00 | 1447.658 | 2.15 | 0.97 | 118.00 | 16740.068 | 2.15 | 0.97 | 118.00 | 18201.921 | 2.15 | 0.97 |
| 119.00 | 1401.907 | 2.15 | 0.97 | 119.00 | 16657.196 | 2.15 | 0.97 | 119.00 | 18149.546 | 2.15 | 0.97 |
| 120.00 | 1357.308 | 2.15 | 0.97 | 120.00 | 16657.195 | 2.14 | 0.97 | 120.00 | 18098.044 | 2.15 | 0.97 |
| 240.00 | 1357.308 | 2.15 | 0.97 | 240.00 | 16657.195 | 2.14 | 0.97 | 240.00 | 18098.044 | 2.15 | 0.97 |
| 241.00 | 1401.907 | 2.15 | 0.97 | 241.00 | 16657.196 | 2.15 | 0.97 | 241.00 | 18149.546 | 2.15 | 0.97 |
| 242.00 | 1447.658 | 2.15 | 0.97 | 242.00 | 16740.068 | 2.15 | 0.97 | 242.00 | 18201.921 | 2.15 | 0.97 |
| 243.00 | 1494.608 | 2.15 | 0.97 | 243.00 | 16783.666 | 2.15 | 0.97 | 243.00 | 18255.544 | 2.15 | 0.97 |
| 244.00 | 1542.751 | 2.15 | 0.97 | 244.00 | 16828.169 | 2.15 | 0.97 | 244.00 | 18309.483 | 2.16 | 0.97 |
| 245.00 | 1592.027 | 2.15 | 0.97 | 245.00 | 16873.213 | 2.15 | 0.97 | 245.00 | 18364.292 | 2.16 | 0.97 |
| 246.00 | 1642.344 | 2.15 | 0.97 | 246.00 | 16918.690 | 2.15 | 0.97 | 246.00 | 18419.866 | 2.16 | 0.97 |
| 247.00 | 1693.663 | 2.16 | 0.97 | 247.00 | 16964.006 | 2.15 | 0.97 | 247.00 | 18476.323 | 2.16 | 0.97 |
| 248.00 | 1746.042 | 2.16 | 0.97 | 248.00 | 17009.449 | 2.16 | 0.97 | 248.00 | 18533.557 | 2.16 | 0.97 |
| 249.00 | 1799.631 | 2.16 | 0.97 | 249.00 | 17055.257 | 2.16 | 0.97 | 249.00 | 18592.029 | 2.16 | 0.97 |
| 250.00 | 1854.623 | 2.16 | 0.97 | 250.00 | 17101.775 | 2.16 | 0.98 | 250.00 | 18652.614 | 2.17 | 0.97 |
| 251.00 | 1911.163 | 2.16 | 0.97 | 251.00 | 17149.599 | 2.16 | 0.98 | 251.00 | 18714.829 | 2.17 | 0.97 |
| 252.00 | 1969.300 | 2.16 | 0.97 | 252.00 | 17199.131 | 2.16 | 0.98 | 252.00 | 18779.400 | 2.17 | 0.97 |
| 253.00 | 2028.996 | 2.16 | 0.97 | 253.00 | 17250.938 | 2.17 | 0.98 | 253.00 | 18845.687 | 2.17 | 0.97 |
| 254.00 | 2090.212 | 2.16 | 0.97 | 254.00 | 17305.303 | 2.17 | 0.98 | 254.00 | 18913.733 | 2.17 | 0.97 |
| 255.00 | 2152.987 | 2.17 | 0.97 | 255.00 | 17362.542 | 2.17 | 0.98 | 255.00 | 18983.691 | 2.17 | 0.97 |
| 256.00 | 2217.459 | 2.17 | 0.97 | 256.00 | 17422.591 | 2.17 | 0.98 | 256.00 | 19054.768 | 2.17 | 0.97 |
| 257.00 | 2283.846 | 2.17 | 0.97 | 257.00 | 17485.261 | 2.17 | 0.98 | 257.00 | 19127.627 | 2.18 | 0.97 |
| 258.00 | 2352.361 | 2.17 | 0.98 | 258.00 | 17550.268 | 2.17 | 0.98 | 258.00 | 19202.012 | 2.18 | 0.97 |
| 259.00 | 2423.161 | 2.17 | 0.98 | 259.00 | 17617.292 | 2.18 | 0.98 | 259.00 | 19278.396 | 2.18 | 0.97 |
| 260.00 | 2496.339 | 2.17 | 0.98 | 260.00 | 17685.851 | 2.18 | 0.98 | 260.00 | 19357.035 | 2.18 | 0.98 |
| 261.00 | 2571.900 | 2.17 | 0.98 | 261.00 | 17755.840 | 2.18 | 0.98 | 261.00 | 19437.270 | 2.18 | 0.98 |
| 262.00 | 2649.803 | 2.17 | 0.98 | 262.00 | 17827.279 | 2.18 | 0.98 | 262.00 | 19519.631 | 2.18 | 0.98 |
| 263.00 | 2730.084 | 2.17 | 0.98 | 263.00 | 17900.510 | 2.18 | 0.98 | 263.00 | 19604.269 | 2.18 | 0.98 |
| 264.00 | 2812.899 | 2.18 | 0.98 | 264.00 | 17976.067 | 2.18 | 0.98 | 264.00 | 19691.391 | 2.19 | 0.98 |
| 265.00 | 2898.501 | 2.18 | 0.98 | 265.00 | 18054.568 | 2.19 | 0.98 | 265.00 | 19781.422 | 2.19 | 0.98 |
| 266.00 | 2987.151 | 2.18 | 0.98 | 266.00 | 18136.341 | 2.19 | 0.98 | 266.00 | 19874.844 | 2.19 | 0.98 |
| 267.00 | 3079.052 | 2.18 | 0.98 | 267.00 | 18221.536 | 2.19 | 0.98 | 267.00 | 19972.465 | 2.19 | 0.98 |
| 268.00 | 3174.339 | 2.18 | 0.98 | 268.00 | 18310.096 | 2.19 | 0.98 | 268.00 | 20073.625 | 2.19 | 0.98 |
| 269.00 | 3273.129 | 2.18 | 0.98 | 269.00 | 18401.998 | 2.19 | 0.98 | 269.00 | 20178.602 | 2.20 | 0.98 |
| 270.00 | 3375.573 | 2.18 | 0.98 | 270.00 | 18497.234 | 2.19 | 0.98 | 270.00 | 20287.879 | 2.20 | 0.98 |